\documentclass[twocolumn]{aastex631}
\usepackage{epsf}
\usepackage{natbib}
\usepackage{graphicx}
\usepackage{float}
\usepackage{amssymb}
\usepackage{enumitem}
\usepackage{verbatim}
\usepackage{stfloats}
\usepackage{bm,bbm,amssymb, amsmath}
\usepackage{enumerate}
\usepackage{url}
\usepackage{soul}
\usepackage{tikz}
\usepackage{forest}
\usepackage{array}
\usepackage{hyperref}
\usepackage{ragged2e}
\usetikzlibrary{arrows.meta,positioning}
\usepackage[section]{placeins}

\usepackage{mathrsfs}
\usepackage{ulem}

\usepackage{tabularx}
\usepackage{hyperref}
\usepackage{xcolor}

\DeclareMathOperator\erf{erf}
\newcommand{\red}{\textcolor{red}}

\newcommand{\brown}[1]{\textcolor{brown}{#1}}
\begin{document}
\title{\Large Map of the Interstellar Continuum FUV Field within 800 Parsecs}
\correspondingauthor{David Lomeli}
\email{dlomeli@utep.edu}

\author[0009-0001-9751-1119]{David Lomeli}
\affil{University of Texas at El Paso, El Paso, TX, 79968, USA}

\author[0000-0003-1572-0505]{Kedron Silsbee}
\affil{University of Texas at El Paso, El Paso, TX, 79968, USA}

\author[0000-0002-1590-1018]{Alexei V. Ivlev}
\affil{Max-Planck-Institut für Extraterrestrische Physik, 85748 Garching, Germany}

\author{Marta Obolentseva}
\affil{Max-Planck-Institut für Extraterrestrische Physik, 85748 Garching, Germany}

\begin{abstract}
The far-ultraviolet (FUV) field, consisting of photons with energies between 6 and 13.6 electron volts, plays an important role in the chemical and thermodynamical evolution of the interstellar medium. FUV continuum radiation is produced by bright stars in the galaxy and absorbed by interstellar dust. We used newly available maps of the dust distribution, in conjunction with catalogs of stellar positions and spectra, to create a map of the stellar FUV field within 800 parsecs of the Earth.  The map shows a total FUV intensity in the solar neighborhood in line with existing measurements (0.7 times the Draine field), but substantial point-to-point fluctuations exist.  Near the Galactic midplane, the FUV intensity decreases with increasing gas density up to a density of approximately 10 hydrogen atoms per cm$^3$, due to absorption of light by dust.  Between 10 and 50 hydrogen atoms per cm$^3$ the UV field increases again, due to a positive correlation between high density gas and young hot stars. We find that the spectrum is generally harder where the UV field is higher, and softer in regions where dust extinction is more important.  The resulting map, as well as our catalog of derived stellar properties, are both available for download for those wishing to study a nearby region of the ISM. 

\end{abstract}

\section{Introduction}
The far ultraviolet (FUV) radiation field, consisting of photons with energies in the range from 6 - 13.6 eV plays an important role in the chemistry of the interstellar medium (ISM). Gas heating in the diffuse ISM is dominated by photoelectric heating from FUV photons \citep{Draine11}, and the photoelectric heating rate has been shown to affect the ratio of the different phases in the interstellar medium \citep{Hill}.  FUV photons also photodissociate and ionize many of the molecules and atoms lying in dense atomic and molecular gas \citep{Tielens85}.  This affects both the chemistry, as well as the thermodynamics, as C$^+$ is one of the dominant coolant species in the cold neutral medium \citep{Wolfire_1, Wolfire_2, Munan}.
\par
Uncertainties in the FUV field have hampered efforts to measure the cosmic ray abundance in local molecular clouds. \citet{Neufeld17} show that the ionization rate inferred from observations of the ${\rm H_3^+}$ molecular ion depends sensitively on the free electron abundance in the gas, which in low-density molecular gas is dominated by photo-ionization of carbon.  This issue was addressed in \citet{Obolentseva24}, who estimated the FUV field at specific points in the Galaxy using a similar method as in the present paper and used this in their chemical modeling. {This work is an extension of the UV field modeling done in \citet{Obolentseva24}.} Imperfect knowledge of the FUV field also complicates measurement of the cosmic ray abundance from ro-vibrational emission of ${\rm H_2}$ molecules, as absorption of UV photons can pump the ${\rm H_2}$ molecule to excited ro-vibrational states \citep{Bialy20}.
\par
There have been previous satellite measurements of the FUV field at the position of Earth. \citet{Henry} present measurements of approximately 1/3 of the sky with a spectrometer aboard the Apollo 17 spacecraft, and demonstrate that these show good agreement with the local field predicted from summing the modeled UV contribution from catalogs of optically bright stars. \citet{Gondhalekar} present measurements from the TD-1 satellite of the UV field from 58,012 stars in four bands centered between 156 and 274 nm.  They estimated also the contribution from ``diffuse Galactic light" to be approximately 10\% of that from direct starlight.  More recent measurements, including those of \citet{Seon11} and \brown{\citet{Murthy10}} have helped us understand the variation in UV across the sky with the use of the SPEAR/FIMS and GALEX missions, respectively.
\par
Aside from satellite measurements, there have also been proposed models of the local FUV field. \citet{Parravano03} modeled the time dependence of the local FUV field and finds temporal variations with amplitudes of a few times due to the evolution of OB associations, supernovae, and runaway OB stars. A separate model, by \citet{Bialy20FUV}, studied the FUV in galactic disks and found dependencies on metallicity, gas density, galactocentric radius, and star-formation rate. Recently, \citet{Bianchi24} modeled the interstellar radiation field at the location of the Sun using all the stars in Gaia DR3, supplemented with bright stars from Hipparcos.  They used the stellar models from \citep{Castelli03} to extrapolate from the optical to the UV. They also made an estimate of the scattered component based on dust maps and IRAS data.
\par
In this work we make a 3D map of the interstellar FUV field in the nearest 800 pc of the ISM by summing the contribution from 862,262 hot stars with distances up to 1.25 kpc. We account for absorption of starlight using the dust extinction map in \citet{Edenhofer24}, but neglect scattering due to its high computational cost. In Section 2, we describe our methodology, including how we compiled our catalog of stellar properties, 
and our treatment of dust extinction.
In Section 3, we present results from the map, focusing on the spectral slope of the FUV field, and how its overall strength varies with gas density, and height above the Galactic midplane. We also compare the FUV spectrum at the location of the Sun with available measurements.
In Section 4, we discuss some limitations of our model, focusing in particular on the incompleteness of our stellar sample, the finite pixel size, and neglect of the scattered radiation field. We present our conclusions in Section \ref{sect:conclusions}.
\par

\section{Methodology}
\label{sect:methodology}
We calculate the FUV flux (or equivalently the energy-density in the FUV field) at each point by directly summing the contribution of a set of 862,262 stars, taking into account the extinction of UV by the intervening dust.  We consider the wavelength range from {$\lambda_{\rm min} = 91.18$ nm  to $\lambda_{\rm max} = 206.64$ nm,} which corresponds to photon energies of 6.0 - 13.5977 eV.  The lower limit of this range is somewhat arbitrary, and we follow here the definition of \citet{Hollenbach99}.  The upper limit is equal to the ionization potential of atomic hydrogen, above which our assumption that extinction is dominated by dust is no longer valid.
\par
From our set of $N$ stars, the energy density at a particular point $j$ is given by 
\begin{equation}
    u_{\rm FUV}^j = \frac{1}{c} \sum_{i=1}^N    \left(\frac{R_i}{D_{ij}}\right)^2 \int_{\lambda_{\rm min}}^{\lambda_{\rm max}} {f_i(\lambda)} e^{-\tau_{ij}(\lambda)} d \lambda
    \label{eq:masterEquation}
\end{equation}
where $R_i$ the radius of star $i$, $D_{ij}$ the distance from star $i$ to point $j$, c the speed of light, ${f_i(\lambda)}$ is the 
flux at the surface of star $i$ and $\tau_{ij}(\lambda)$ the line-of-sight extinction between star $i$ and point $j$ 

\subsection{Stellar Models}
To determine ${f_i(\lambda)}$ used in Equation \eqref{eq:masterEquation}, we used the stellar atmosphere models from \cite{Castelli03}. 
These models are based on a star's log(g) (logarithm of the surface gravity in cgs units), metallicity, and effective temperature $T_{\rm eff}$. 
\par
\begin{figure}[h!]
    \centering
    \includegraphics[width=\linewidth]{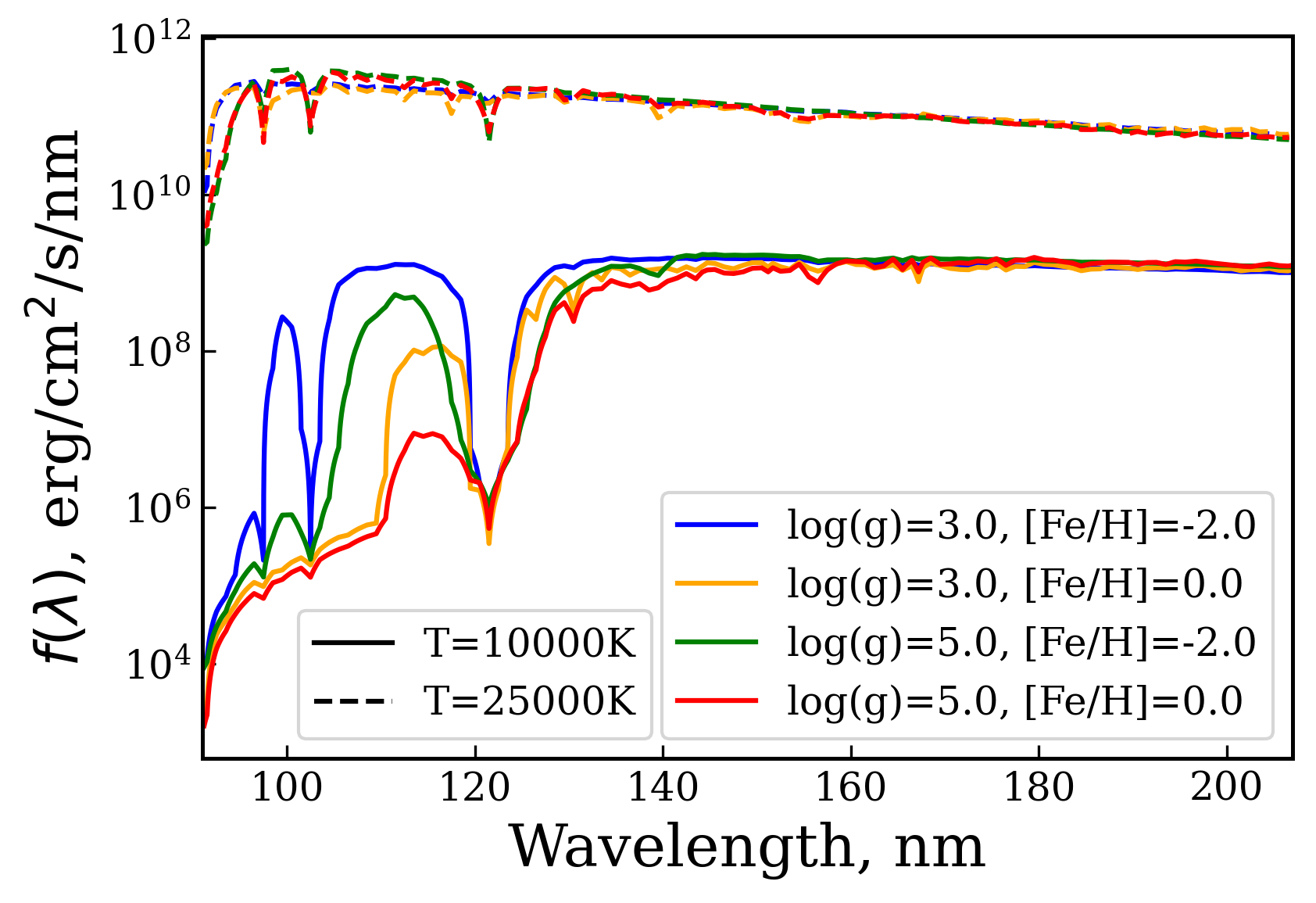}
    \caption{Flux ${f(\lambda)}$ for eight different stellar models from \cite{Castelli03} with varying $T_{\rm eff}$, log(g), and metallicities.  The effective temperature is the main parameter determining the spectrum, but metallicity and log(g) play a role as well, particularly shortward of Lyman ${\alpha}$ (121.6 nm).}
    \label{fig: spectra_combs}
\end{figure}

Figure \ref{fig: spectra_combs} shows a representative sample of these models.  The majority of the difference between models is attributable to the effective temperature, with metallicity and log(g) playing a secondary role.  For that reason, an estimate of the effective temperature is a necessary condition for inclusion of a star in our catalog.  If log(g) and metallicity are available, then we use those values.  For stars for which no log(g) or metallicity value was available, the mean of the rest of the catalog was assumed. These values are 4.05 dex and -0.12 dex, respectively.  For the highest temperature stars, models with log(g) = 4 were not available and, in this case, we used log(g) = 4.5 or 5, whenever applicable.  We used trilinear interpolation between the discrete temperature, log(g), and metallicity models given in \citet{Castelli03} to determine ${f(\lambda)}$ for stars of arbitrary parameters. 

\subsection{Determination of Stellar Parameters}
At a minimum, in order to include a star as a source in our model, we need to know its 3D position in space, radius and temperature. In addition, if available, we use the star's log(g) and metallicity. To obtain these parameters, we use several catalogs. The first is the PASTEL catalog, \citep{pastel}, which includes estimates of the effective temperature for 31,401 stars based on high-resolution spectroscopy. 
Second, we use the Tycho-2 Spectral Type Catalog \citep{Tycho2SP} which is a subset of Tycho-2 containing 351,864 stars which were cross-matched from Tycho-2 to several catalogs containing spectral types.  This catalog allows us to assign values of the effective temperature $T_{\rm eff}$ and Tycho $V_T$ band magnitude for the stars in our catalog. 
We also make use of the Tycho Double Star Catalog \citep{tdsc}. This catalog includes stars in binary systems with small angular separations, mostly within 0.3 to 1 arcseconds.
We then use the catalog in \citet{Starhorse}, 
which provides effective temperatures, masses, and surface gravities for 362 million stars, based on data from Gaia EDR3, cross-matched with several other photometric catalogs.  This catalog is made using the StarHorse code \citep{Queiroz18}, and we refer to it as ``StarHorse-EDR3" in this work.
Lastly, we supplemented our catalog by adding distance estimates for stars from Hipparcos \citep{hip2} in cases where distance estimates from Gaia were not available. There are two instances in which the distances were manually changed. Epsilon and Iota Orionis distances were changed to 361 pc as per \citet{Oplistilova23}.

\par

\begin{figure*}[p]
\raggedright
\begin{forest}
for tree={
    grow'=east,
    draw,
    rounded corners=2pt,
    align=left,
    text width=1.8cm,
    minimum height=0.6cm,
    font=\footnotesize\sffamily,
    parent anchor=east,
    child anchor=west,
    anchor = west,
    edge path={
        \noexpand\path[\forestoption{edge}]
        (!u.parent anchor) -- +(5pt,0) |- (.child anchor)\forestoption{edge label};
    },
    l sep=8mm,
    s sep=3mm,
    edge={-latex, semithick},
    tier/.wrap pgfmath arg={tier #1}{level()},
}
[$V_T$
    [Pastel?
        [Eq. \ref{eq:V_to_VT}, edge label={node[midway,above,font=\tiny\bfseries]{Y}}]
        [Tycho-2?, edge label={node[midway,below,font=\tiny\bfseries]{N}}
            [Tycho-2, edge label={node[midway,above,font=\tiny\bfseries]{Y}}]
            [TDSC?, edge label={node[midway,below,font=\tiny\bfseries]{N}}
                [TDSC, edge label={node[midway,above,font=\tiny\bfseries]{Y}}]
                [GAIA?, edge label={node[midway,below,font=\tiny\bfseries]{N}}
                    [Eq. \ref{eq:VTfromGAIA}, edge label={node[midway,above,font=\tiny\bfseries]{Y}}]
                    [NaN, edge label={node[midway,below,font=\tiny\bfseries]{N}}]
                ]
            ]
        ]
    ]
]
\end{forest}

\begin{forest}
for tree={
    grow'=east,
    draw,
    rounded corners=2pt,
    align=left,
    text width=1.8cm,
    minimum height=0.6cm,
    font=\footnotesize\sffamily,
    parent anchor=east,
    child anchor=west,
    anchor=west,
    edge path={
        \noexpand\path[\forestoption{edge}]
        (!u.parent anchor) -- +(5pt,0) |- (.child anchor)\forestoption{edge label};
    },
    l sep=8mm,
    s sep=3mm,
    edge={-latex, semithick},
    tier/.wrap pgfmath arg={tier #1}{level()},
}
[$T_{\rm eff}$
    [Pastel?
        [Pastel, edge label={node[midway,above,font=\tiny\bfseries]{Y}}]
        [Tycho-2?, edge label={node[midway,below,font=\tiny\bfseries]{N}}
            [Tycho-2, edge label={node[midway,above,font=\tiny\bfseries]{Y}}]
            [Starhorse?, edge label={node[midway,below,font=\tiny\bfseries]{N}}
                [Starhorse, edge label={node[midway,above,font=\tiny\bfseries]{Y}}]
                [NaN, edge label={node[midway,below,font=\tiny\bfseries]{N}}
                ]
            ]
        ]
    ]
]
\end{forest}

\begin{forest}
for tree={
    grow'=east,
    draw,
    rounded corners=2pt,
    align=left,
    text width=1.8cm,
    minimum height=0.6cm,
    font=\footnotesize\sffamily,
    parent anchor=east,
    child anchor=west,
    anchor=west,
    edge path={
        \noexpand\path[\forestoption{edge}]
        (!u.parent anchor) -- +(5pt,0) |- (.child anchor)\forestoption{edge label};
    },
    l sep=8mm,
    s sep=3mm,
    edge={-latex, semithick},
    tier/.wrap pgfmath arg={tier #1}{level()},
}
[$\rm log(g)$,
    [Pastel?
        [Pastel, edge label={node[midway,above,font=\tiny\bfseries]{Y}}]
        [Starhorse?, edge label={node[midway,below,font=\tiny\bfseries]{N}}
            [Starhorse, edge label={node[midway,above,font=\tiny\bfseries]{Y}}]
            [NaN, edge label={node[midway,below,font=\tiny\bfseries]{N}}
            ]
        ]
    ]
]
\end{forest}

\begin{forest}
for tree={
    grow'=east,
    draw,
    rounded corners=2pt,
    align=left,
    text width=1.8cm,
    minimum height=0.6cm,
    font=\footnotesize\sffamily,
    parent anchor=east,
    child anchor=west,
    anchor = west,
    edge path={
        \noexpand\path[\forestoption{edge}]
        (!u.parent anchor) -- +(5pt,0) |- (.child anchor)\forestoption{edge label};
    },
    l sep=8mm,
    s sep=3mm,
    edge={-latex, semithick},
    tier/.wrap pgfmath arg={tier #1}{level()},
}
[$A_V$
    [Starhorse?
        [Starhorse, edge label={node[midway,above,font=\tiny\bfseries]{Y}}]
        [Dustmap, edge label={node[midway,below,font=\tiny\bfseries]{N}}
            ]
        ]
    ]
\end{forest}

\begin{forest}
for tree={
    grow'=east,
    draw,
    rounded corners=2pt,
    align=left,
    text width=1.8cm,
    minimum height=0.6cm,
    font=\footnotesize\sffamily,
    parent anchor=east,
    child anchor=west,
    anchor=west,
    edge path={
        \noexpand\path[\forestoption{edge}]
        (!u.parent anchor) -- +(5pt,0) |- (.child anchor)\forestoption{edge label};
    },
    l sep=8mm,
    s sep=3mm,
    edge={-latex, semithick},
    tier/.wrap pgfmath arg={tier #1}{level()},
}
[Fe/H
    [Pastel?
        [Pastel, edge label={node[midway,above,font=\tiny\bfseries]{Y}}]
        [Starhorse?, edge label={node[midway,below,font=\tiny\bfseries]{N}}
            [Starhorse, edge label={node[midway,above,font=\tiny\bfseries]{Y}}]
            [NaN, edge label={node[midway,below,font=\tiny\bfseries]{N}}
            ]
        ]
    ]
]
\end{forest}

\begin{forest}
for tree={
    grow'=east,
    draw,
    rounded corners=2pt,
    align=left,
    text width=1.8cm,
    minimum height=0.6cm,
    font=\footnotesize\sffamily,
    parent anchor=east,
    child anchor=west,
    edge path={
        \noexpand\path[\forestoption{edge}]
        (!u.parent anchor) -- +(5pt,0) |- (.child anchor)\forestoption{edge label};
    },
    l sep=8mm,
    s sep=3mm,
    edge={-latex, semithick},
    tier/.wrap pgfmath arg={tier #1}{level()},
}
[Distance
    [Starhorse?
        [Starhorse, edge label={node[midway,above,font=\tiny\bfseries]{Y}}]
        [GAIA?, edge label={node[midway,below,font=\tiny\bfseries]{N}}
            [GAIA, edge label={node[midway,above,font=\tiny\bfseries]{Y}}]
            [Hipparcos?, edge label={node[midway,below,font=\tiny\bfseries]{N}}
                [Hipparcos, edge label={node[midway,above,font=\tiny\bfseries]{Y}}]
                [NaN, edge label={node[midway,below,font=\tiny\bfseries]{N}}
                ]
            ]
        ]
    ]
]
\end{forest}

\begin{forest}
for tree={
    grow'=east,
    draw,
    rounded corners=2pt,
    align=left,
    text width=1.8cm,
    minimum height=0.6cm,
    font=\footnotesize\sffamily,
    parent anchor=east,
    child anchor=west,
    anchor = west,
    edge path={
        \noexpand\path[\forestoption{edge}]
        (!u.parent anchor) -- +(5pt,0) |- (.child anchor)\forestoption{edge label};
    },
    l sep=8mm,
    s sep=3mm,
    edge={-latex, semithick},
    tier/.wrap pgfmath arg={tier #1}{level()},
}
[Angular \\Coordinates
    [Starhorse?
        [Starhorse, edge label={node[midway,above,font=\tiny\bfseries]{Y}}]
        [Pastel?, edge label={node[midway,below,font=\tiny\bfseries]{N}}
            [Pastel, edge label={node[midway,above,font=\tiny\bfseries]{Y}}]
            [Tycho-2?, edge label={node[midway,below,font=\tiny\bfseries]{N}}
                [Tycho-2, edge label={node[midway,above,font=\tiny\bfseries]{Y}}]
                [Hipparcos?, edge label={node[midway,below,font=\tiny\bfseries]{N}}
                    [Hipparcos, edge label={node[midway,above,font=\tiny\bfseries]{Y}}]
                    [NaN, edge label={node[midway,below,font=\tiny\bfseries]{N}}]
                ]
            ]
        ]
    ]
]
\end{forest}

\begin{forest}
for tree={
    grow'=east,
    draw,
    rounded corners=2pt,
    align=left,
    text width=1.8cm,
    minimum height=0.6cm,
    font=\footnotesize\sffamily,
    parent anchor=east,
    child anchor=west,
    anchor = west,
    edge path={
        \noexpand\path[\forestoption{edge}]
        (!u.parent anchor) -- +(5pt,0) |- (.child anchor)\forestoption{edge label};
    },
    l sep=8mm,
    s sep=3mm,
    edge={-latex, semithick},
    tier/.wrap pgfmath arg={tier #1}{level()},
}
[Mass
    [Starhorse?
        [Starhorse, edge label={node[midway,above,font=\tiny\bfseries]{Y}}]
        [NaN, edge label={node[midway,below,font=\tiny\bfseries]{N}}
            ]
        ]
    ]
]
\end{forest}

\begin{forest}
for tree={
    grow'=east,
    draw,
    rounded corners=2pt,
    align=left,
    text width=1.8cm,
    minimum height=0.6cm,
    font=\footnotesize\sffamily,
    parent anchor=east,
    child anchor=west,
    anchor=west,
    edge path={
        \noexpand\path[\forestoption{edge}]
        (!u.parent anchor) -- +(5pt,0) |- (.child anchor)\forestoption{edge label};
    },
    l sep=8mm,
    s sep=3mm,
    edge={-latex, semithick},
    tier/.wrap pgfmath arg={tier #1}{level()},
}
[Radius
    [$V_T$?
        [Eq. \ref{eq:magnitudeToRadius}, edge label={node[midway,above,font=\tiny\bfseries]{Y}}]
        [Starhorse?, edge label={node[midway,below,font=\tiny\bfseries]{N}}
            [Eq. \ref{eq:RfromMass}, edge label={node[midway,above,font=\tiny\bfseries]{Y}}]
            [NaN, edge label={node[midway,below,font=\tiny\bfseries]{N}}
            ]
        ]
    ]
]
\end{forest}

\caption{Decision trees showing the hierarchical procedure for acquiring the parameters in our catalog. In coordinates, if a star does not have a distance from GAIA or Hipparcos and the star does not appear in Starhorse-EDR3, then its coordinates are not calculated and star is not included. If a radius is not able to be calculated for a particular star, then that star is not included either.} 
\label{Decision_trees}
\end{figure*}
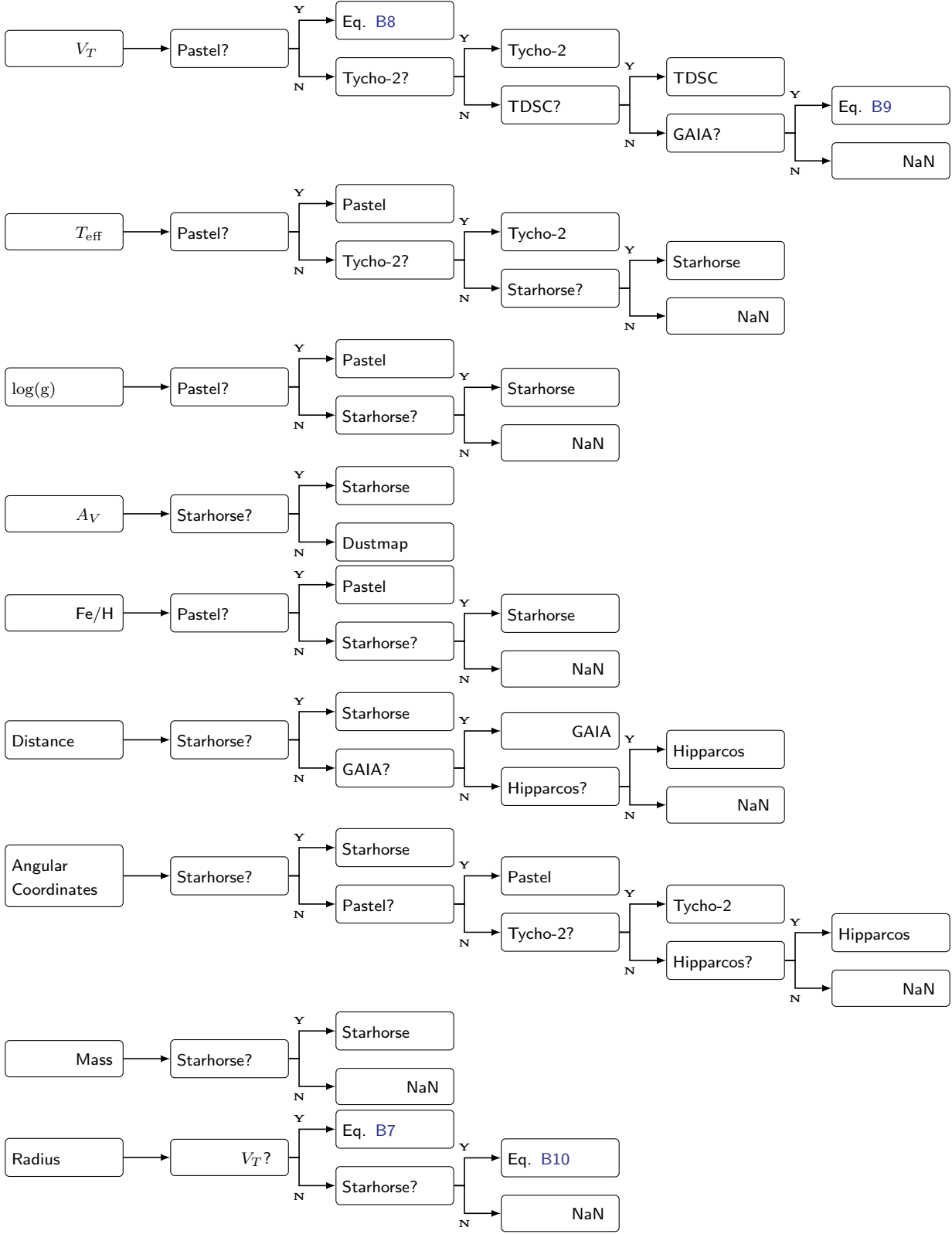

 Figure \ref{Decision_trees} illustrates the procedure by which we selected the parameters for our catalog.
In the case of coordinates, Starhorse-EDR3 gives 3D Galactic Cartesian coordinates and require no further transformation. In all other cases, the distance is acquired from GAIA or Hipparcos and the SkyCoord package from Astropy 
is utilized to transform angular coordinates to Galactic Cartesian coordinates. The two methods of deriving stellar radius from observables referred to in Figure \ref{Decision_trees} are described in Appendix \ref{Stellar radius}.

{The priority given to the different catalogs is different for each variable. That said, we often prioritize PASTEL as it consists of stars measured with high-resolution spectroscopy.}

\indent {Note that our catalog was constructed slightly differently from the stellar catalog from \citet{Obolentseva24}. In that work, $V_T$ values were not calculated using equations \ref{eq:V_to_VT} or \ref{eq:VTfromGAIA}. Therefore, the stellar radii were calculated using Eq. \ref{eq:RfromMass} in cases where the $V_T$ magnitude was not directly available. Distances were acquired from GAIA whenever possible, otherwise they were acquired from the Hipparcos 2 catalog.}

\par

\subsection{Extinction by Dust}
\label{sect:dustExtinction}
To account for extinction, we used the dust map from \citet{Edenhofer24}, interpolated onto a Cartesian grid composed of cubical cells with side-length 5 parsecs. This dust map gives the visual extinction per unit distance based on analysis of the reddening of 54 million stars with known distance. In constructing the map, they assumed the dust extinction curve from \citet{Zhang}. The map covers a spherical region around the Earth with radius of 1.25 kiloparsecs. The map contains a spherical hole of radius 69 pc centered at the Earth. We neglect any extinction within this hole. {To minimize inaccuracies from neglect of more distant stars, we limit the range of our FUV map to 800 parsecs.  this choice is justified in Section \ref{sect:distanceLimit}.} To evaluate the extinction from a star to a point $x$, we employ a very simple algorithm for computational efficiency.  We sampled the differential extinction at 100 evenly spaced points along a line connecting $x$ to the star, and calculated the integrated visual extinction $\tau_V$ using the trapezoid rule. The numerical errors stemming from only using 100 points increase with increasing gas density, but are typically a few percent or smaller. We collected a sample of 800 random points in our volume, and an additional 200 randomly selected points with gas density $n_H > 10$ cm$^{-3}$.  For these points, we calculated the energy density using our standard algorithm in which extinction was evaluated at 100 points along the line of sight, and then again using 1000 points along each line of sight. We found that the average magnitude of the fractional difference between the energy density calculated in these two different ways was 0.09\% for points with $n_H < 10$ cm$^{-3}$, and 1.54\% for points with $n_H > 10$ cm$^{-3}$.

\par
To calculate $\tau(\lambda)$ from $\tau_V$, we use the extinction curve from \citet{Cardelli}, assuming a spatially invariant value of $R_V = 3.1$.  Because the extinction curve in \citet{Cardelli} only extends to 12.4 eV, we use a quadratic polynomial to extrapolate their curve to 13.6 eV.
\par
{We employed the following procedure to accelerate the computation of Equation \eqref{eq:masterEquation}}.  We divided the extinction curve and stellar spectra into six sub-intervals.  We then define $u_k$ as the integrated energy density over the respective wavelength range.  We normalize our UV field to Equation 11 from \citet{Draine}.  This is a fitting function based on several theoretical and observational studies from the 1960s and 1970s.  This integrates to a total energy density $U_D = 8.92 \times 10^{-14}$ erg cm$^{-3}$ in photons between 6 and 13.6 eV.  We show these wavelength intervals in Table \ref{tab:uv_bins}, as well as the energy density $u_{D_k}$ in the Draine field in each interval.  We define $G_{D_k} = u_k/u_{D_k}$, and $G_D = \sum_{k=1}^6 u_k/\sum_{k = 1}^6 u_{D_k}$.

For each interval, we calculate the ratio of the average extinction over that interval to $A_V$
\begin{equation}
    \bar A_k = \frac{\int_{\lambda_{\rm k,min}}^{\lambda_{\rm k, max}} A_\lambda/A_V d\lambda}{\lambda_{\rm k, max} - \lambda_{\rm k,min}}.
\end{equation}
We then approximate the total UV field energy density at point $j$ as 
\begin{equation}
u_{\rm FUV}^j = \frac{1}{c} \sum_{i=1}^N \sum_{k=1}^{6} \left(\frac{R_i}{D_{ij}}\right)^2 F_{ik} e^{-\bar{A}_k \tau_V^{ij}}
\label{eq:masterWithBins}
\end{equation}

Where ${F_{ik}}$ is the flux from star $i$ {integrated} over 
wavelength interval k, $D_{ij}$ the distance between star $i$ and point $j$ and $\tau_V^{ij}$ the V-band optical depth between star $i$ and point $j$.

\begin{table}[h!]
    \centering
    \setlength{\tabcolsep}{3pt} 
    \begin{tabular}{|c|c|c|c|}
        \hline
        {Bin} & {Wavelength} & {Energy} & {`Draine' Field} \\
        & {Range (nm)} & {Range (eV)} & ($\times 10^{-14}$ erg/cm$^3$) \\
        \hline
        $u_1$ & 91.18 -- 110.42 & 13.6 -- 11.23 & 0.95\\
        $u_2$ & 110.42 -- 129.67 & 11.23 -- 9.56 & 1.22\\
        $u_3$ & 129.67 -- 148.91 & 9.56 -- 8.33 & 1.55\\
        $u_4$ & 148.91 -- 168.15 & 8.33 -- 7.37 & 1.90\\
        $u_5$ & 168.15 -- 187.40 & 7.37 -- 6.61 & 2.02\\
        $u_6$ & 187.40 -- 206.64 & 6.61 -- 6.00 & 1.29\\
        \hline
    \end{tabular}
    \caption{
    Wavelength and energy ranges for the six FUV bins and the values $u_{D_k}$ of the integrated `Draine' field within each bin.  }
    \label{tab:uv_bins}
\end{table}

\subsection{Catalog Constraints}
\begin{figure}[h!]
    \centering
    \includegraphics[width=\linewidth]{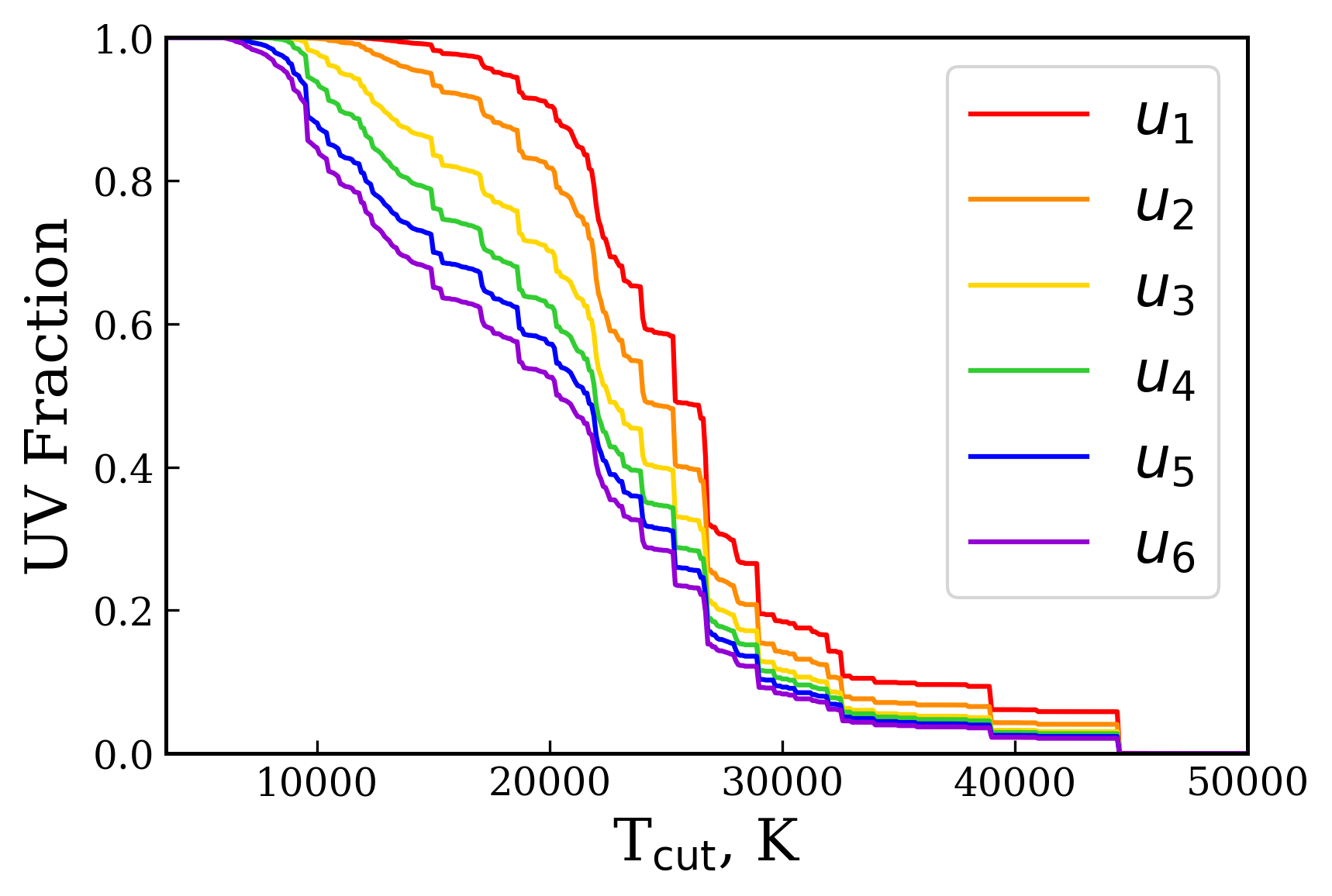} 
    \caption{Fraction of the energy density at the location of the Sun contributed by stars with temperature greater than $T_{\rm cut}$ as a function of $T_{\rm cut}$.  Different color curves correspond to the six wavelength bins enumerated in Table \ref{tab:uv_bins}.  The curves for each bin are normalized separately. The discontinuities are due to the discrete contributions from individual stars.}
    \label{fig: u_T_cut}
\end{figure}
In constructing our final set of 862,262 stars, we 
included only stars within 1.25 kpc of the Sun, with effective temperatures $\geq$ 7,000 Kelvin.  The distance cut is necessary because the dust map in \citet{Edenhofer24} does not extend beyond this radius, and thus we have no way of evaluating $\tau_V$ in Equation \eqref{eq:masterWithBins} for more distant stars.  As shown below, the temperature cut has a negligible impact on our results.
\par
Figure \ref{fig: u_T_cut} shows the fraction of the energy density at the location of the Sun coming from those stars with temperatures greater than a cut-off temperature $T_{\rm cut}$ in each of the wavelength bins shown in Table \ref{tab:uv_bins}.  For this analysis we used all stars in the previously mentioned catalogs that are within 1,250 parsecs.  We find that stars with effective temperatures less than 7,000 Kelvin only contribute 0.17\% of the FUV energy density at the location of the Sun in total, and no more than 1.3\% even in the longest wavelength bin (to which low-temperature stars have the highest fractional contribution).  
We do not include the Sun itself in this analysis. 

\section{Results}
Using the methodology in Section \ref{sect:methodology} we constructed a map of the UV field in a spherical volume with radius of 800 parsecs surrounding the Sun. The FUV map produced in this work, as well as our catalog of stellar properties are available at Harvard Dataverse \cite{Lomeli26_Data}.

\subsection{Spatial Distribution of the FUV Field}
\begin{figure*}[p]
    \centering
    \includegraphics[width = \linewidth]{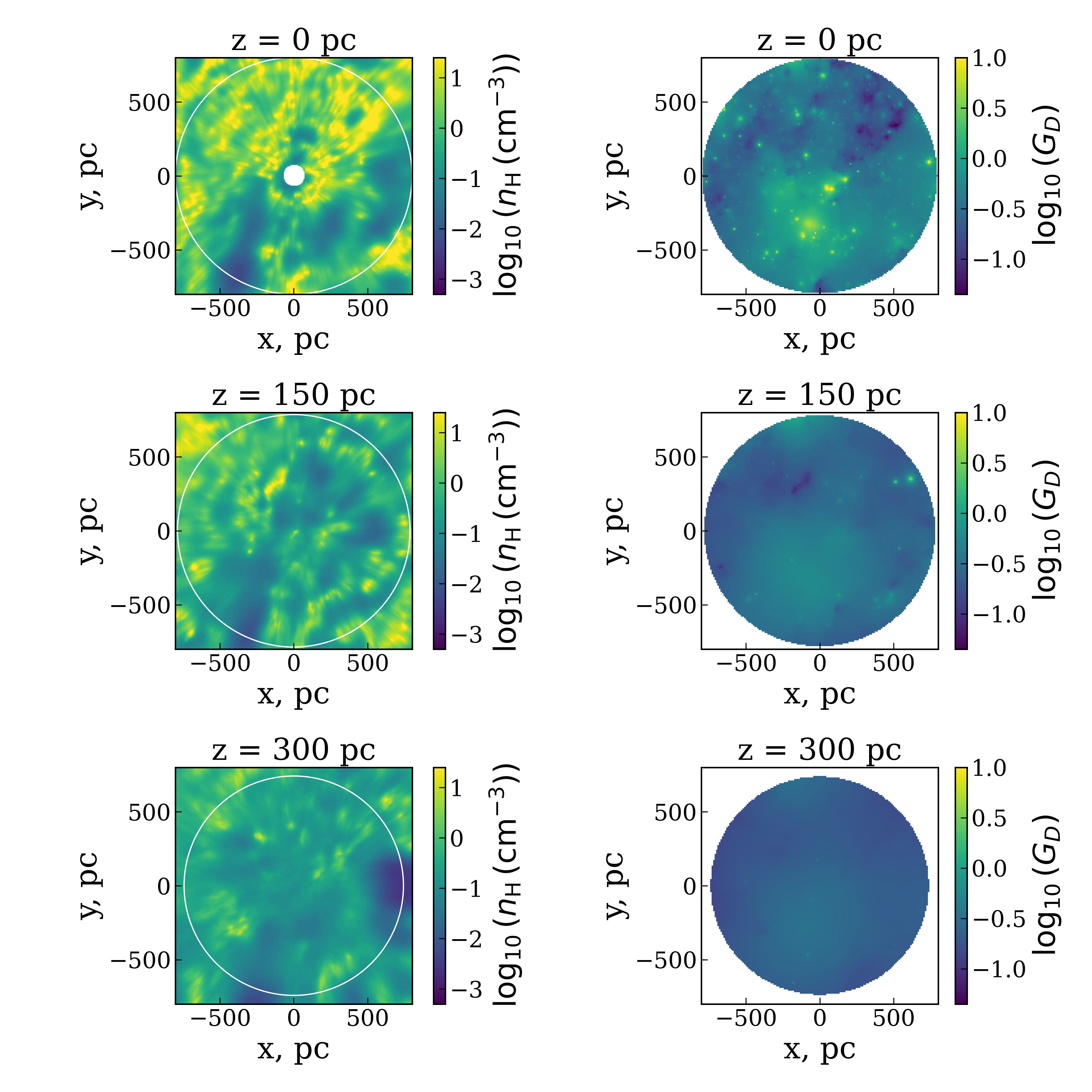}
    \caption{Slices of the dust map from \citet{Edenhofer24} (left) and our FUV map (right) at heights of 0, 150, and 300 pc, where z = 0 corresponds to the height of the Sun in the Milky Way. We have converted the visual extinction density ${\rm E_V}$ reported in \citet{Edenhofer24} to a gas density, using the relation $\frac{n_{\rm H}}{\rm cm^{-3}} =716 {\rm E_V}$, where ${\rm E_V}$ is measured in units of ${\rm A_V}$/pc \citep{Guver09}.  The hole in the middle of the top left panel corresponds to the region that is excluded from the dust-map.  Circles are added onto the dust-map plots to show the area spanned by the UV plots on the right. We see as expected that the gas density is higher near $z = 0$.  The plots on the right show that the UV field is also generally higher near $z = 0$, as well as less uniform, due to the higher abundance of hot stars and obscuring dust clouds. 
    }
    \label{fig:dust_uv_0}
\end{figure*}

Figure \ref{fig:dust_uv_0} shows three slices from the dust-map of \citet{Edenhofer24} and from our corresponding UV field maps. We have converted the visual extinction density ${\rm E_V}$ reported in \citet{Edenhofer24} to a gas density, using the relation $\frac{n_{\rm H}}{\rm cm^{-3}} =716 {\rm E_V}$, where ${\rm E_V}$ is measured in units of ${\rm A_V}$/pc, {not to be confused with the commonly used color excess E(B-V). This relation follows from the relation ${N_H(\rm cm^{-2}) = 2.21\times 10^{21} A_V (\rm mag)}$ from \citet{Guver09}}. We notice two particular patterns in the slices.  First, there are small circular regions of anomalously high UV field. These are regions in which the UV field is dominated by a single star, or compact star cluster.  In some cases, an individual pixel is anomalously bright in comparison to the neighboring pixels.  This is due to a star of moderate UV brightness that is unusually close to the center of a grid cell.  We discuss this case further in Section \ref{sect:mapSmoothing}.

In addition, there are also irregularly shaped regions of reduced FUV intensity, particularly in the slice at $z = 0$.  These appear to be spatially co-located with regions of high extinction density, suggesting that they are caused by absorption of UV photons by dust.
\par
\begin{figure}
    \centering
    \includegraphics[width = \linewidth]{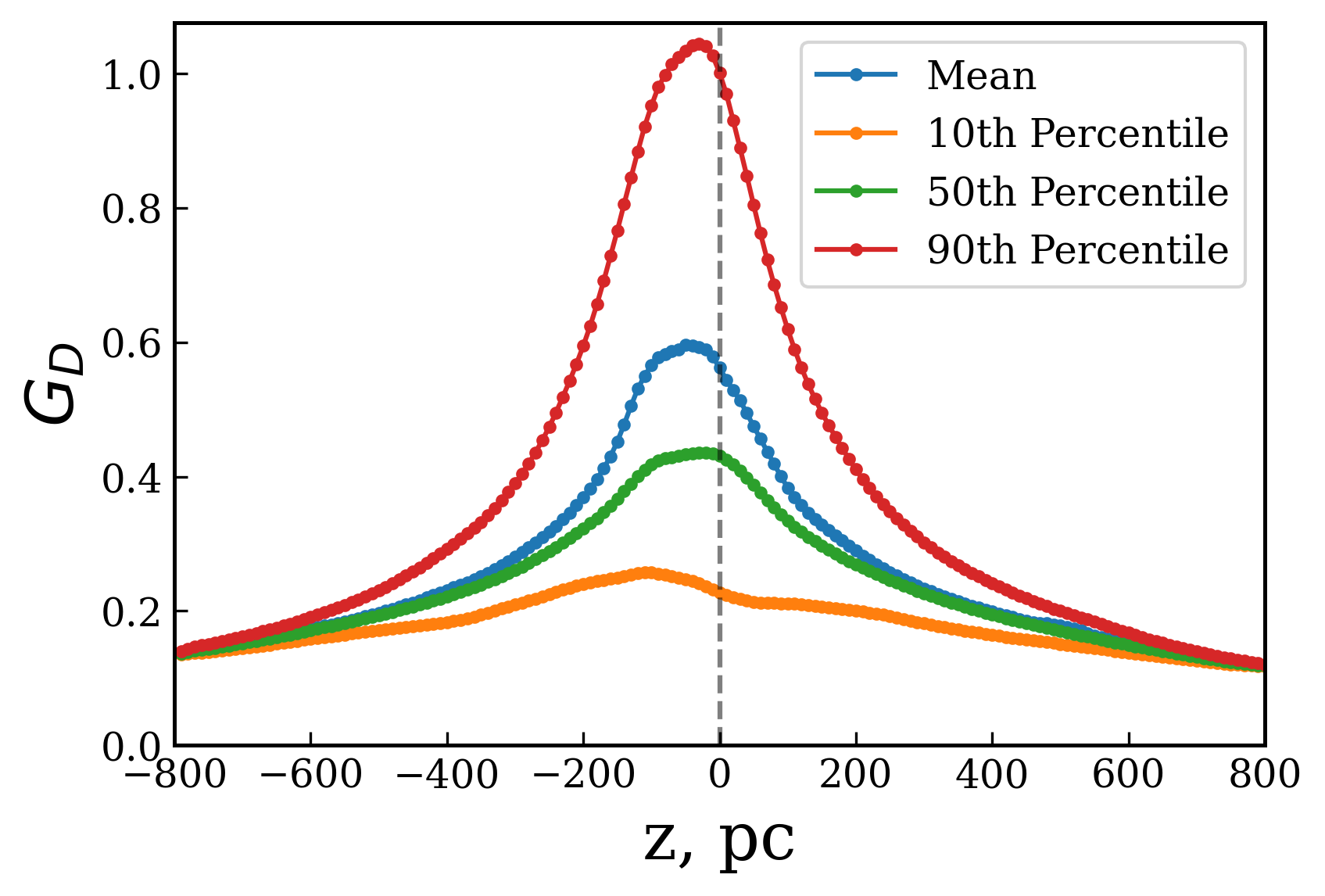}
    \caption{Statistics of the FUV field as a function of Galactic $z$ for points within our map. The peak of the UV field is offset from $z = 0$ because the Sun is slightly above the Galactic midplane. The  value of each curve at $z$ represents statistics of the set of points $x'$, $y'$, $z'$ in our map with $|z' - z| \leq 30$ pc.
    }
    \label{fig:quantiles_vs_z}
\end{figure}

Figure \ref{fig:quantiles_vs_z} shows the vertical variation of the mean energy density. The distribution is smooth, peaking at approximately $z = -30$ pc, consistent with other evidence that the Sun is slightly above the Galactic midplane \citep{Karim17}. The mean energy density decreases by a factor of $\sim$3 at $|z|$ = 800 pc compared to near the Galactic midplane.  As discussed in Section \ref{sect:distanceLimit}, some of this reduction is due to the increasing importance of the incompleteness of our sample with increasing distance from the Galactic midplane.

\subsection{FUV Spectrum and Energy Density at the Sun}
As a test of our method, we calculated the spectrum of the interstellar UV field at the location of the Sun. 
In this case, instead of just six energy bins, we broke the spectrum into 11,547 energy bins, corresponding to wavelength intervals of 0.01 nm.  We found an integrated FUV energy density of $u_{FUV} = 6.56\times10^{-14}$ erg/cm$^3$ corresponding to $G_D = 0.74$. Equation \eqref{eq:masterWithBins} gives a value of $6.64\times10^{-14}$ erg/cm$^3$ at the location of the Sun; implying a 1.7\% error from the coarse binning.
\par
\begin{figure}
    \centering
    \includegraphics[width=\linewidth]{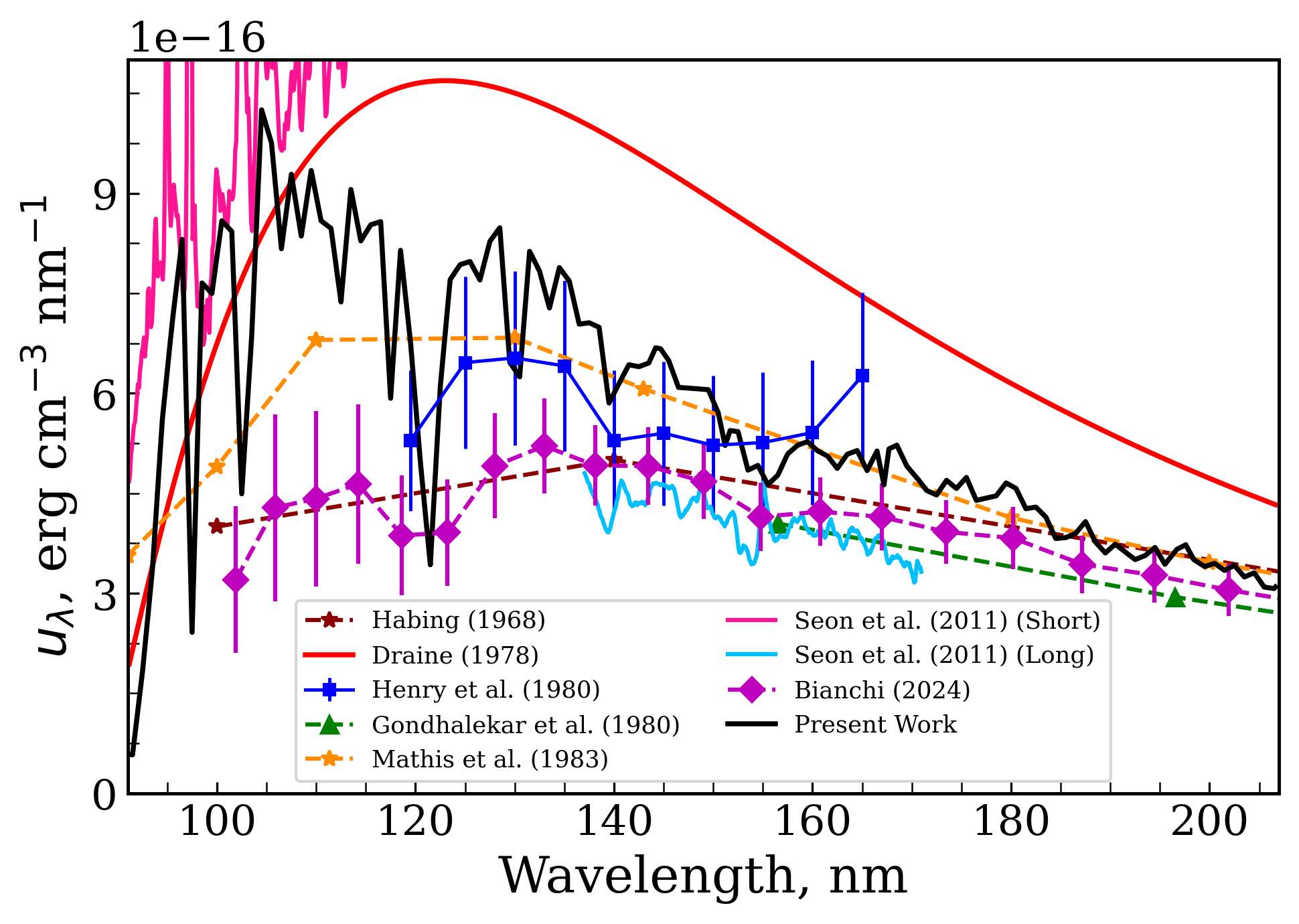}
    \caption{Derived spectrum in the FUV range at the location of the Sun compared to observational values from \cite{Gondhalekar}, \citet{Henry} and \citet{Seon11}. (Note that we have corrected the FIMS data for their incomplete sky coverage; see text for details).  In addition, we show the model FUV spectra from {\citet{Habing}}, \citet{Draine}, {\citet{Mathis1983}}, and \citet{Bianchi24}.}
    \label{fig: nu_U_UV_spectra}
\end{figure}

Figure~\ref{fig: nu_U_UV_spectra} shows our derived FUV spectrum. We compare our spectra with observations from the Apollo 17 mission \cite{Henry}, the ESRO TD-1 satellite \cite{Gondhalekar}, data from the FIMS instrument \citet{Seon11}, modeled LISRF from \citet{Bianchi24}, {\citet{Habing}, \citet{Mathis1983}}, and the widely adopted ``Draine field" from \cite{Draine}.  We note that the FIMS instrument did not survey the entire sky. In an attempt to correct for this, the curve we plot is the average flux per solid angle measured by FIMS, multiplied by $4 \pi f_{\rm sky}/f_{\rm model}$.  Here, $f_{\rm sky}$ is the fraction of the sky that FIMS observed in the relevant wavelength interval. $f_{\rm model}$ is fraction of the UV field in our model at the location of the Sun coming from the part of the sky measured with FIMS. $f_{\rm sky}$ = 0.74 and 0.82, and $f_{\rm model}$ = 0.60 and 0.68 for the long and short wavelength ranges respectively.  This curve is our estimate of what FIMS would have measured over the whole sky, if the ratio of flux measured by FIMS to that from our model were independent of position. 
 \par
 The data from \cite{Henry} are slightly higher than that of \cite{Gondhalekar} in the region of wavelength overlap.  {Part, but not all of this difference comes from the fact that we plot the ``direct starlight component" for \citet{Gondhalekar}, whereas the measurements from \citet{Henry} also include scattered radiation.} 
 The FIMS data match our spectrum reasonably well at short wavelengths except for prominent emission lines (notably Lyman $\beta$ and Lyman $\gamma$, which appear as absorption features in the stellar spectra, and for which we do not attempt to  model the radiative transfer). On the longer wavelength end, there is a lower amount of radiation, compared with both our model and other observations. 
\indent We also plot {four} other models for the local UV field.  First, we show the widely-used ``Draine Field" \citep{Draine} discussed in Section \ref{sect:dustExtinction}. We also show the {total} LISRF fit model from \citet{Bianchi24}, which used all the stars in the Gaia DR3 catalog supplemented with additional bright stars from Hipparcos to model the direct LISRF. They estimate the diffuse UV radiation based on published correlations with 100 $\mu m$ emission. Their spectrum is generally below ours. We hypothesize that this discrepancy results from differences between our methods of inferring stellar properties. We also include the comparison with the ISRF from \citet{Mathis1983}, which is in fair agreement with our calculations. This field was produced by fitting a three-component field to the measured total ISRF at the time and also produced estimates at different galactocentric distances by varying those components. We use their ISRF model at 10 kpc from the Galactic center, as they indicate they believed that to be representative of the solar neighborhood. We note however, that the Mathis field is intended to represent the total ISRF, whereas ours is only the direct component.  If we included the scattered light as well, the discrepancy would be somewhat larger (see Section \ref{sect:scattering}). Lastly, we show the `Habing' field (Table 8 from \citet{Habing}), which was created from a model distribution of OB stars and dust. This field is generally below ours as well.
{We further compare our results with these previously published values in Table \ref{tab:fuv_comparison}. We integrate the spectra in Figure \ref{fig: nu_U_UV_spectra} and divide by our result over the corresponding wavelength range denoted in Column 2. The discrete-valued fields were interpolated.}
\par
We do not mean to imply that the agreement with the local measurements is typical of the accuracy of our model.  We expect our method to do better at the location of the Sun than at other places, because at the location of the Sun, the contribution from stars whose radii we calculate using Equation \eqref{eq:magnitudeToRadius} is largely insensitive to uncertainties in the distance since the distance is chosen so that the model optical flux from the star is equal to the observed value. However, the agreement with observations suggests that our relation between observed stellar properties and UV luminosity (which has no tunable parameters) is on average accurate to within the uncertainties in our knowledge of the local field.

\subsection{Low Resolution Spectrum of the FUV field}

\begin{figure*}
    \centering
    \includegraphics[width=\textwidth]{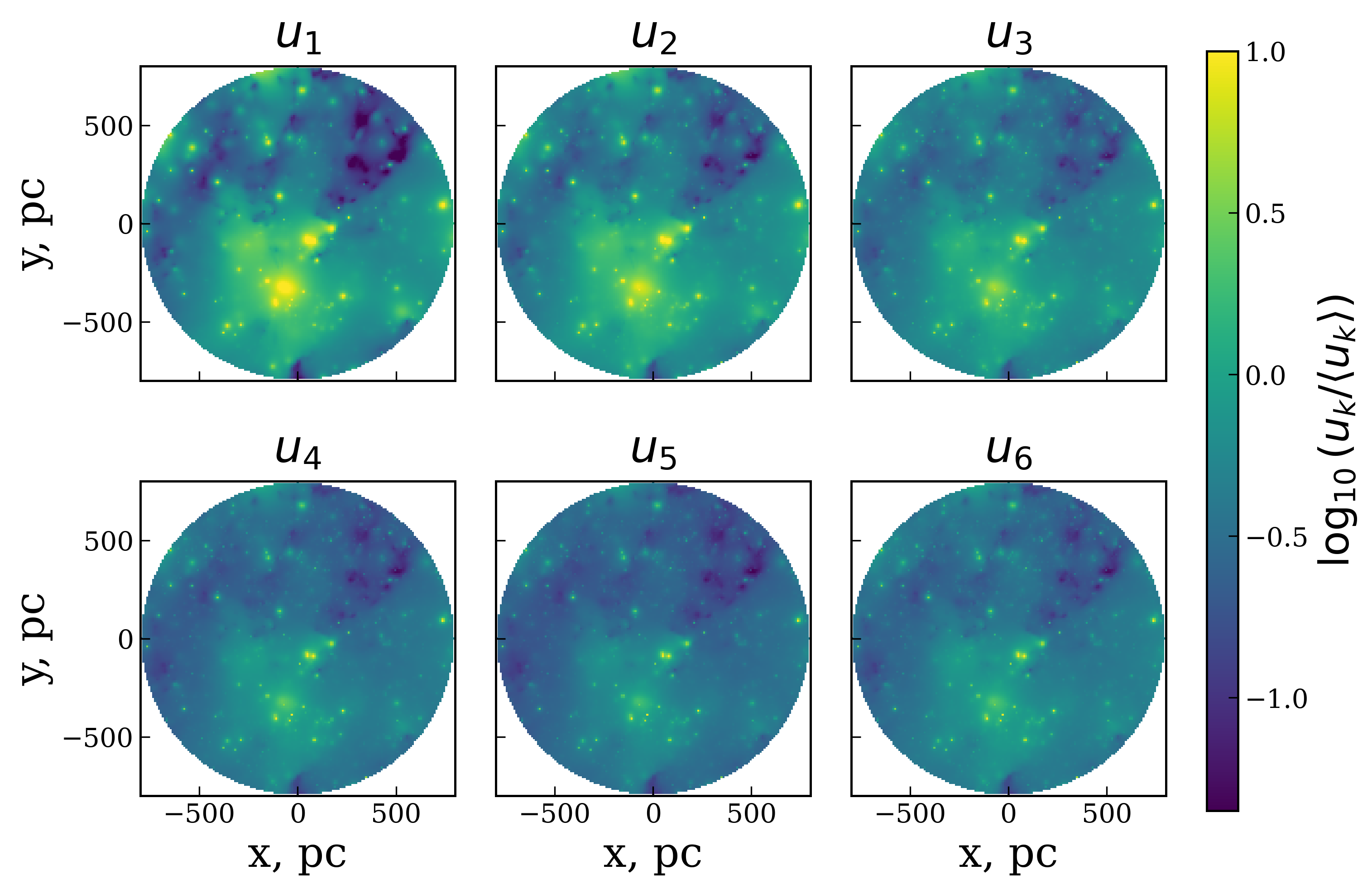}
    \caption{Map of the UV field in the plane $z = 0$ in each wavelength bin normalized to the {average energy density for each bin at z = 0.}. We observe that the UV field exhibits higher spatial variability in the shorter wavelength bins. 
    }
    \label{fig:uv_spectra}
\end{figure*}

In this section, we examine how the distribution of the energy density in the 6 wavelength bins shown in Table \ref{tab:uv_bins} varies throughout the map. Figure \ref{fig:uv_spectra} shows the FUV field at $z = 0$ for each of these bins. Each subplot of Figure \ref{fig:uv_spectra} shows a broadly similar pattern, however the variations in UV field are more pronounced in the shortest wavelength bin ($u_1$) compared to the longest ($u_6$). This is to be expected, since the most luminous stars are typically very hot and rare, and attenuation of starlight due to dust is stronger at shorter wavelengths.

\begin{table*}
    \centering
    \setlength{\tabcolsep}{3pt}
    \begin{tabular}{|l|c|c|c|c|}
        \hline
        {Dataset} &
        {Wavelength Range} &
        {Published $\int u_\lambda\, d\lambda$} &
        {Present Work $\int u_\lambda\, d\lambda$} &
        {Ratio} \\
        & {(nm)} & {(erg cm$^{-3}$)} & {(erg cm$^{-3}$)} & \\
        \hline
        \cite{Habing}           & 100.00 -- 206.64 & $4.49\times10^{-14}$ & $6.12\times10^{-14}$ & 0.73 \\
        \cite{Draine}           & 91.18 -- 206.64  & $8.92\times10^{-14}$ & $6.56\times10^{-14}$ & 1.36 \\
        \cite{Henry}            & 119.50 -- 165.00 & $2.62\times10^{-14}$ & $2.84\times10^{-14}$ & 0.92 \\
        \cite{Gondhalekar}      & 156.50 -- 206.64 & $1.67\times10^{-14}$ & $2.12\times10^{-14}$ & 0.79 \\
        \cite{Mathis1983}       & 91.18 -- 206.64  & $6.01\times10^{-14}$ & $6.56\times10^{-14}$ & 0.92 \\
        \cite{Seon11} (Short)   & 91.18 -- 114.95  & $3.19\times10^{-14}$ & $1.69\times10^{-14}$ & 1.89 \\
        \cite{Seon11} (Long)    & 137.05 -- 170.95 & $1.37\times10^{-14}$ & $1.88\times10^{-14}$ & 0.73 \\
        \cite{Bianchi24} (2024) & 101.92 -- 206.64 & $4.30\times10^{-14}$ & $5.96\times10^{-14}$ & 0.72 \\
        \hline
    \end{tabular}
    \caption{Comparison of integrated spectra from previous measurements and models shown in Figure \ref{fig: nu_U_UV_spectra} with the present work. Columns 3 and 4 correspond to the integrated spectra over the common wavelength range between those measurements and models and the present work. Data points are interpolated when necessary.}
    \label{tab:fuv_comparison}
\end{table*}

To further investigate the changes in the spectrum, we first define the ratio $\mathscr{R} = {u_6}/{u_1}$. 
At the Sun, $\mathscr{R}_0 = 0.54$.

\begin{figure*}
    \centering
    \includegraphics[width=\textwidth]{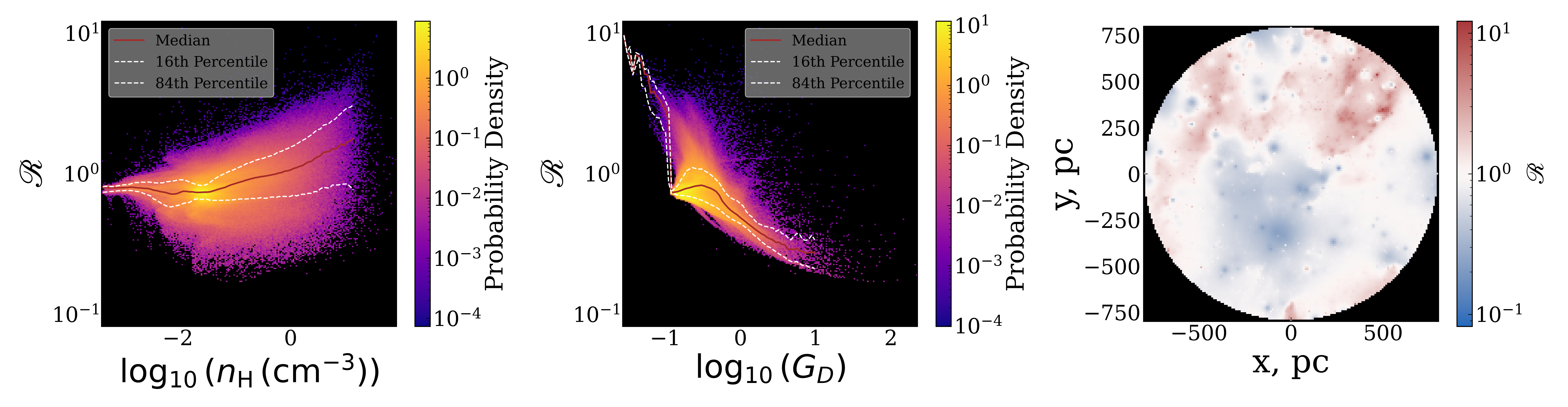}
    \caption{Left: 2D histogram of $\mathscr{R}$ versus gas density. 
    Probability density is expressed per unit $\log_{10} \rm {n_H}$ per unit $\mathscr{R}$. Middle: 2D histogram of $\mathscr{R}$ versus energy density coming from our map. Right: Slice at z = 0 of $\mathscr{R}$. 
    }
    \label{fig:alpha}
\end{figure*}

Figure \ref{fig:alpha} shows how $\mathscr{R}$ varies with gas density (left panel), overall strength of FUV field (middle panel) and position in the plane $z = 0$ (right panel).

 The left panel of Figure \ref{fig:alpha} shows a weak positive correlation between $\mathscr{R}$ and gas density $n_H$. This trend is consistent with the fact that the extinction is greater at shorter wavelengths. The middle panel shows a dramatic decrease in $\mathscr{R}$ with increasing FUV field.  We attribute this to the fact that regions with strong UV field are located near clusters of young hot stars.  Since only the hottest stars in our sample contribute significantly to $u_1$ (see Figure \ref{fig: u_T_cut}), it is not unexpected that locations near very hot stars will have both a strong FUV field, and a hard FUV spectrum.  The right panel of Figure \ref{fig:alpha} shows the spatial variation of the spectral index in the plane $z = 0$.  Comparing with the top right panel of Figure \ref{fig:dust_uv_0}, we note that the broad features associated with dust extinction in Figure \ref{fig:dust_uv_0} show up as high positive values of $\mathscr{R}$, whereas {most of the} the narrow axisymmetric features (corresponding to individual hot stars or tightly-bound clusters) have values of $\mathscr{R}$ {below the median for this slice}, indicating a harder spectrum. 

 \subsection{Correlation With Gas Density}

\begin{figure}
    \centering
    \includegraphics[width=\linewidth]{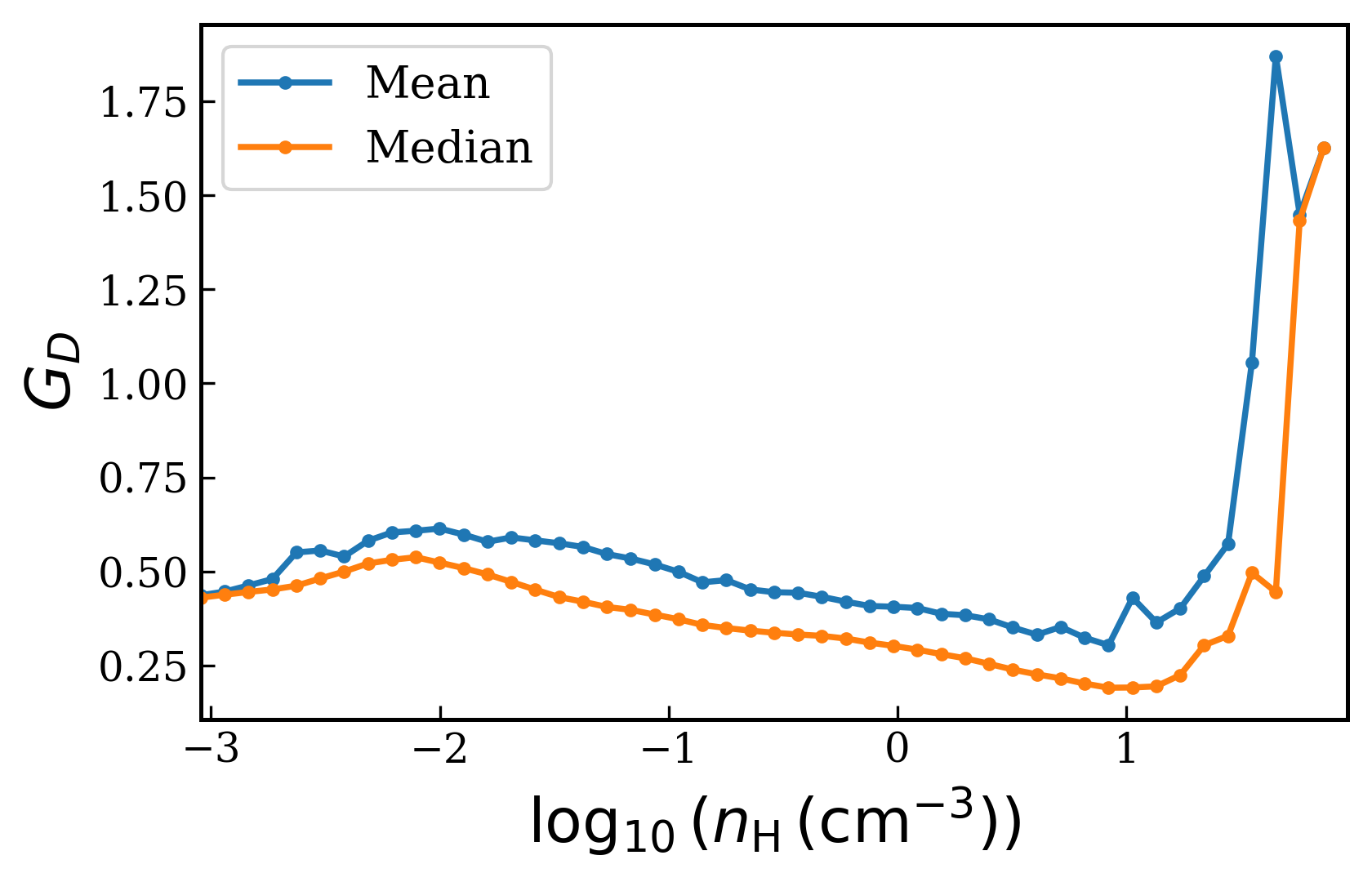}
    \caption{Relation between the mean and median of the normalized total energy density of all regions with average gas density ${n_{\rm H}}$ within $|z| \leq 150$ pc of the Sun.
    }
    \label{fig:u_vs_nh}
\end{figure}

In Figure \ref{fig:dust_uv_0}, we notice an anti-correlation between the gas density and the UV field strength. This anti-correlation is more apparent in regions closer to the midplane where dust extinction generally plays a greater role. 
Figure \ref{fig:u_vs_nh} shows the mean and median UV field as a function of gas density within 150 pc of the midplane. 
Between $-2 <\log_{10}(n_{\rm H}) < 1$, we observe a downward trend where areas of increasingly higher dust density tend to have lower values of total energy density due to higher extinction. For $\log_{10}(n_H) > 1$, the trend reverses, perhaps indicating that hot stars are preferentially near dense gas.  However, since we have very few grid points with such high average gas densities, so we do not make strong claims regarding this density regime. We observe that below a density of $10^{-2}$ cm$^{-3}$, the UV field again decreases with decreasing gas density.  We do not posit an explanation for this trend, but do note that gas with density below $10^{-2}$ cm$^{-3}$ does not contribute substantially to the extinction of starlight.  For example, gas density of $10^{-2}$ cm$^{-3}$ multiplied by a path length equal to the diameter of our map (1,600 pc) yields only a visual extinction of 0.02. {We hypothesize that the map in \citet{Edenhofer24} may not be able to reliably characterize very diffuse gas due to its negligible contribution to the extinction. We did not attempt to quantify this uncertainty.}

\section{Discussion}
\label{sect:disc}
\subsection{Effect of our Distance-Limited Stellar Sample.}
\label{sect:distanceLimit}

\begin{figure}
    \centering
    \includegraphics[width=\linewidth]{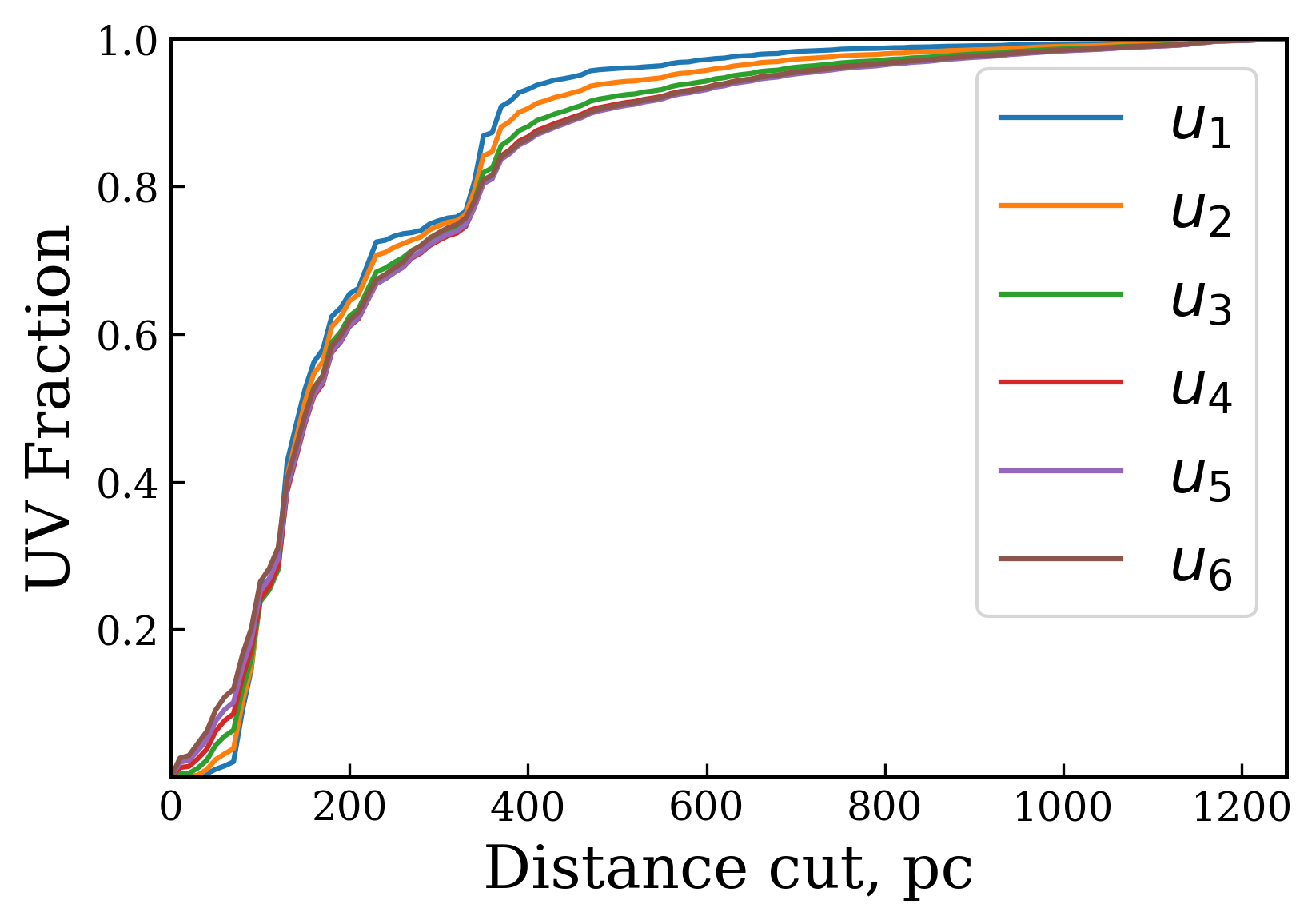}
    \caption{Ratio of the energy density at the Sun attributable to stars within distance $d$ to that attributable to stars within 1250 pc, as a function of $d$.  Different color curves correspond to the different wavelength ranges shown in Table \ref{tab:uv_bins}.  Sharp jumps in the plot are caused by individual UV-luminous stars.}
    \label{fig:f_v_d}  
\end{figure}

To help us estimate the effect of not including stars that lie beyond the dust map, we plot in Figure \ref{fig:f_v_d} the fraction of the UV energy density at the location of the Sun that is attributable to stars lying within a given distance, as a function of that distance.  Over the entire FUV band, stars with distances between 0 and 450 pc contribute 90.3\% as much energy density as those with distances between 0 and 1,250 pc. We observe a weak trend that a higher fraction of shorter-wavelength light is attributable to stars within a given distance cut.  We hypothesize that this is due to the increasing extinction at shorter wavelengths.  
We assume that stars beyond 1,250 pc contribute negligibly to the flux at the Sun, and therefore that only about 10\% of the UV field at the location of the Sun comes from stars further than 450 pc.

The 800 parsec radius of our map ensures that each point in our map lies more than 450 pc from the edge of the map in \citet{Edenhofer24}. If the result in Figure \ref{fig:f_v_d} is typical —that is if 90\% of the UV field at a given point is attributable to stars within 450 pc, then the errors in our map due to non-inclusion of stars beyond the edge of the map in \citet{Edenhofer24} should be lower than 10\%.  We caution, as discussed below, that the errors due to omission of more distant stars are almost certainly larger outside of the Galactic midplane.
\par
We also compared these results to a toy model of the Galaxy in which the population of hot stars has a Gaussian distribution centered on $z = 0$ with standard height $\sigma_*$, and the differential optical depth in the UV (taken as wavelength-independent in this model) has a Gaussian distribution centered on $z = 0$ with mid-plane value $A_0$ and standard height $\sigma_{\rm ext}$.  We ignore any radial variation in the differential extinction or stellar density, and assume it to extend to infinity in all directions.  We show in Appendix \ref{sect:appendix} that under these assumptions, at distance $z$ above the Galactic plane, the UV flux arising from stars out to radial distance $R$ is proportional to  

\begin{equation}
\label{eq:Fcont}
\begin{aligned}
    &F \propto \int_{r = 0}^R r dr \\
     & \times \int_{z' = -\sqrt{R^2 - r^2}+ z}^{\sqrt{R^2 - r^2}+ z} \exp{\left(\frac{-z'^2}{2 \sigma_*^2}\right)} \frac{\exp{\left[-\tau_{\rm UV}(r, z', z)\right]}}{r^2 + (z'-z)^2} dz'
\end{aligned}
\end{equation}
where 
\begin{flalign}
    \label{eq:taucont}
    &\tau_{\rm UV}(r, z', z) =  \frac{\sqrt{(z'-z)^2 + r^2}}{|z'-z|}\sqrt{\frac{\pi}{2}} \tau_0 \sigma_{\rm ext} \nonumber &&\\
    & \times \left|\erf{\left(\frac{z'}{\sigma_{\rm ext} \sqrt{2}}\right)} - \erf{\left(\frac{z}{\sigma_{\rm ext} \sqrt{2}}\right)}\right| &&
\end{flalign}
\begin{figure}
    \centering
    \includegraphics[width=\linewidth]{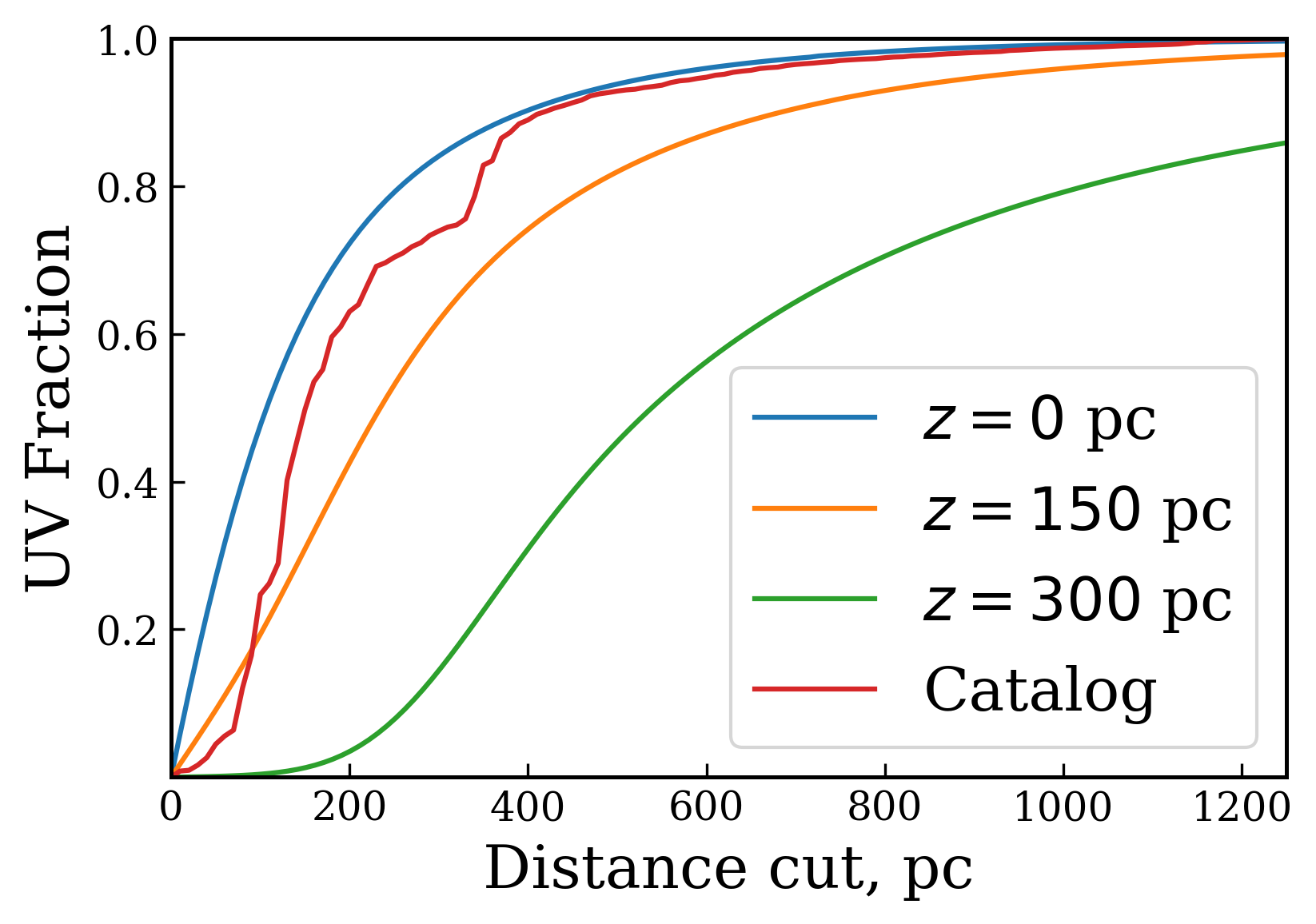}
    \caption{Fraction of flux coming from stars within distance $R$ as a function of $R$.  This was calculated with the approximate continuous model described by Equations \eqref{eq:Fcont} and \eqref{eq:taucont}. The red line shows the fraction of the total FUV flux from stars in our catalog based on a distance cut from the Sun. 
    }
    \label{fig:continuousApproximation}
\end{figure}
We fit Gaussians to the average differential extinction, and the average FUV luminosity per unit volume by stars in our catalog as a function of Galactic $z$ to estimate $\tau_0 = 3.35 \times 10^{-3}$ pc$^{-1}$, $\sigma_{\rm ext} = 107$ pc, and $\sigma_* = 83$ pc.  
We approximate $\tau_{\rm UV} = 3.27 \tau_V$, which is the average ratio over the entire FUV wavelength range. We used Equations \eqref{eq:Fcont} and \eqref{eq:taucont} to calculate the UV flux from stars out to distance $R$, normalized by the flux in the limit as $R \rightarrow \infty$.  We did this for $z = 0$, $z = 150$ pc, and $z = 300$ pc.   The results are shown in Figure \ref{fig:continuousApproximation}.  Near the mid-plane, 99\% of the flux comes from stars within 800 parsecs.  At 150 and 300 pc above the midplane, this number decreases to 96\% and 76\% respectively, due to the lower differential extinction and lack of nearby stars contributing to the UV field.  We note that this is a very approximate model, and the assumption of a smooth dust density distribution {\it overestimates} the role of dust attenuation, as compared to a more realistic clumpy ISM structure. That said, particularly at distances greater than 400 pc, the $z = 0$ curve agrees remarkably well with the results at the location of the Sun from our catalog. Based on these considerations, we caution that the UV field is likely underestimated by up to 10 percent in the Galactic midplane near the edges of the map, and likely by much more in regions more than a dust scale-height above the midplane.

\subsection{Effect of Neglecting Scattering}
\label{sect:scattering}

 The map from \citet{Edenhofer24} is a map of differential {\it extinction}, which includes both scattering and absorption opacity.  Therefore, by using $\tau$ derived from this map in Equation \eqref{eq:masterWithBins}, 
 we are ignoring the contribution of any scattered UV photons to the local radiation field.  We can estimate the effect of neglecting this factor by making a separate map in which the extinction opacity is replaced with the absorption opacity. To implement this, we assume the dust grains to have wavelength-dependent albedo $\varpi(\lambda)$ given in \citet{Hensley23} (extrapolated with a quadratic polynomial between 100 and 91.18 nm). Over the range of $\lambda$ considered in this paper, $\varpi$ varies between 0.15 and 0.37. We then multiply $\bar{A_k}$ in Equation \eqref{eq:masterWithBins} by $1-\bar{\varpi}$, where $\bar\varpi$ is the average $\varpi$ over the wavelength range corresponding to bin $k$, to obtain the optical depth due just to absorption.  
 This effectively adds back the scattered photons, but assumes that the scattering angle is negligibly small.  We do not expect this procedure to correctly account for the spatial distribution of the scattered light (since the scattering angle need not necessarily be small), but comparing the difference between this map and that created with our standard procedure gives a rough estimate of the expected degree of error due to scattering. 
 \par
\begin{figure}
    \centering
    \includegraphics[width=\linewidth]{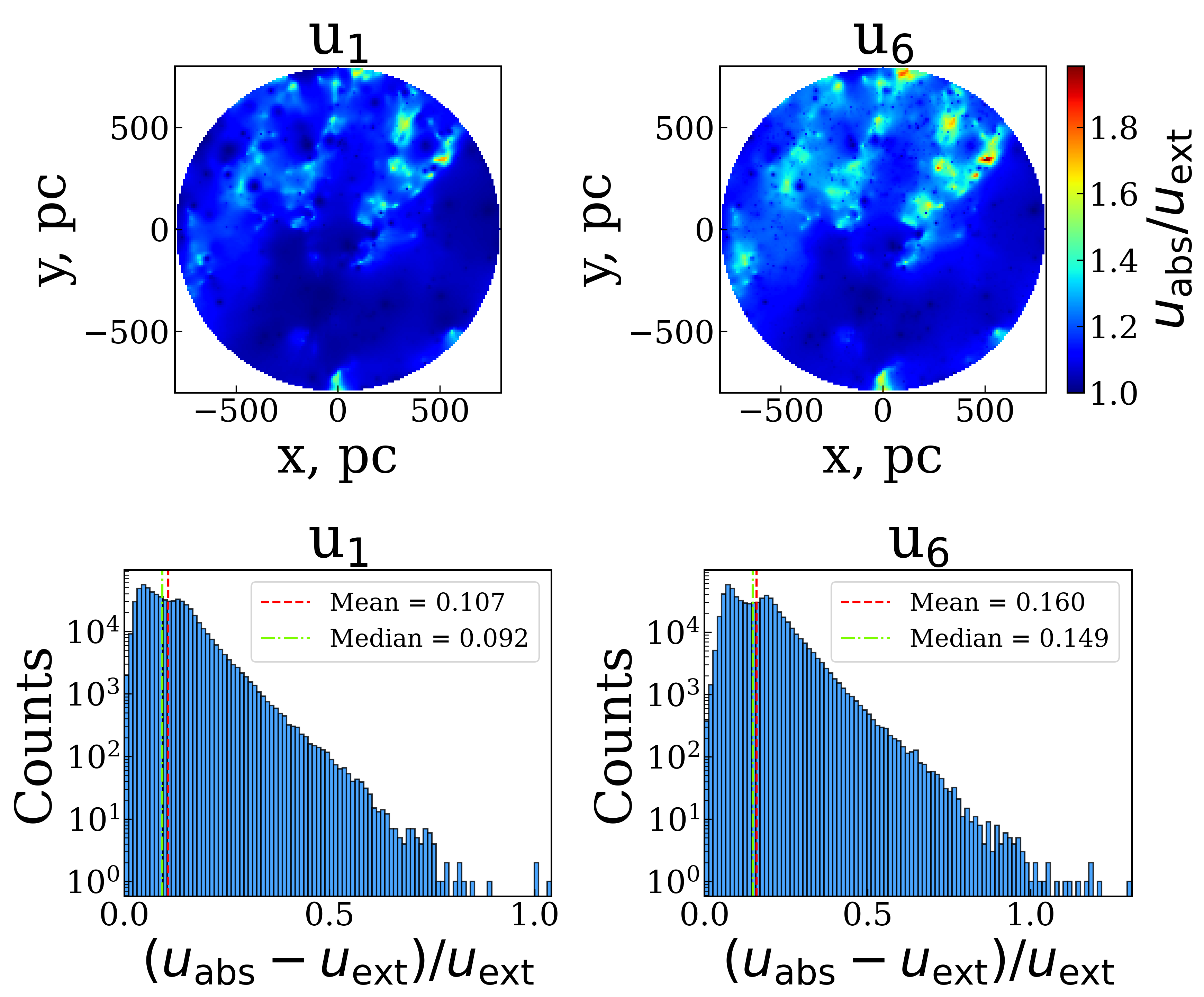}
    \caption{The top panels show a map of the ratio between energy density considering only absorption, $u_{\rm abs}$, and energy density considering absorption and scattering, $u_{\rm ext}$, for $u_1$ and $u_6$ in the z = 0 plane.  The bottom panels show a histogram of the distribution of the fractional difference for each wavelength bin from $|z| \leq 150$ pc. The mean and median of that distribution are also shown. }
    \label{fig:u_abs}
\end{figure}

 The results of this procedure are shown in Figure \ref{fig:u_abs}.  
 The top panels of Figure \ref{fig:u_abs} show maps of the ratio of the UV field with just the absorption opacity considered to that with both absorption and scattering.  The left panel corresponds to the lowest-wavelength bin ($u_1$), and the right panel corresponds to the highest-wavelength bin ($u_6$). 
 By comparison with the top panel of Figure \ref{fig:dust_uv_0}, we see that regions of high $u_{\rm abs}$/$u_{\rm ext}$ are spatially associated with regions of low UV field, and high dust density.  This confirms our expectation that scattering is playing a larger role in those regions.  Note also that individual bright stars can be seen as small  blue circular regions (where the ratio is nearly unity), since the UV field in those regions is dominated by that one star, to which the extinction is negligible.
 \par
 The bottom panels of Figure \ref{fig:u_abs} show the distribution of the fractional increase in the UV field when only the absorption opacity is considered.  The mean and median value for each wavelength bin are shown on the plots. While this test is by no means quantitatively rigorous, we note that our results are roughly consistent with the ratio of diffuse galactic light to direct starlight reported in \citet{Gondhalekar}.
 \par
\begin{figure}
    \centering
    \includegraphics[width=\linewidth]{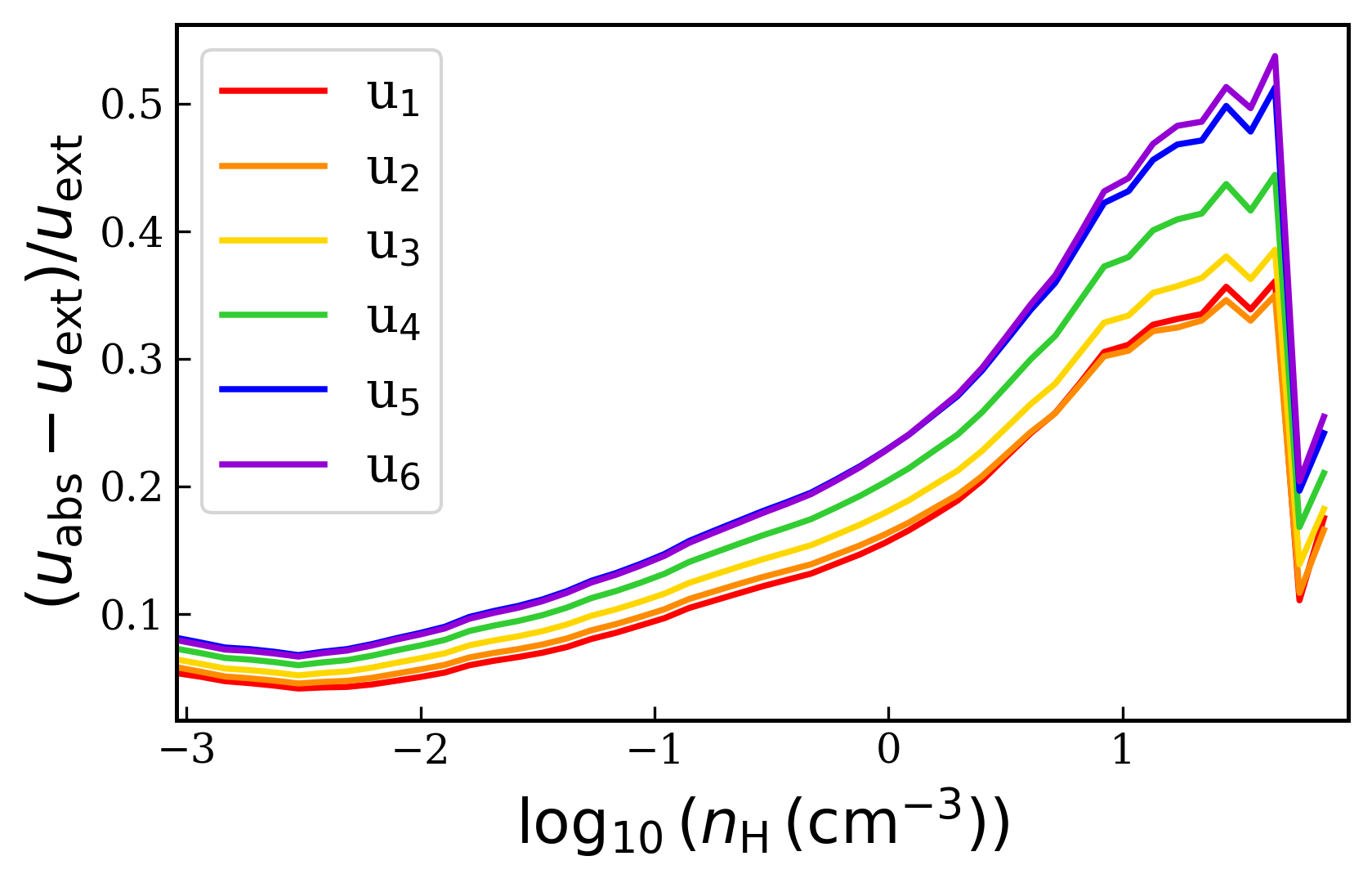}
    \caption{Fractional difference between the average energy density considering only absorption and that considering absorption and scattering, as a function of gas density. Each curve corresponds to one wavelength bin, and the average is taken over all voxels in the map with $|z| \leq 150$ pc.}
    \label{fig:u_abs_n_h}
\end{figure}

 Figure \ref{fig:u_abs_n_h} shows the average fractional increase in the UV field from ignoring scattering opacity as a function of gas density.  As expected, the increase is highest at the highest densities, due to the larger impact of dust shielding at those densities.  

\subsubsection{Smoothing of the map}
\label{sect:mapSmoothing}
\begin{figure*}
    \includegraphics[width=0.98\textwidth]{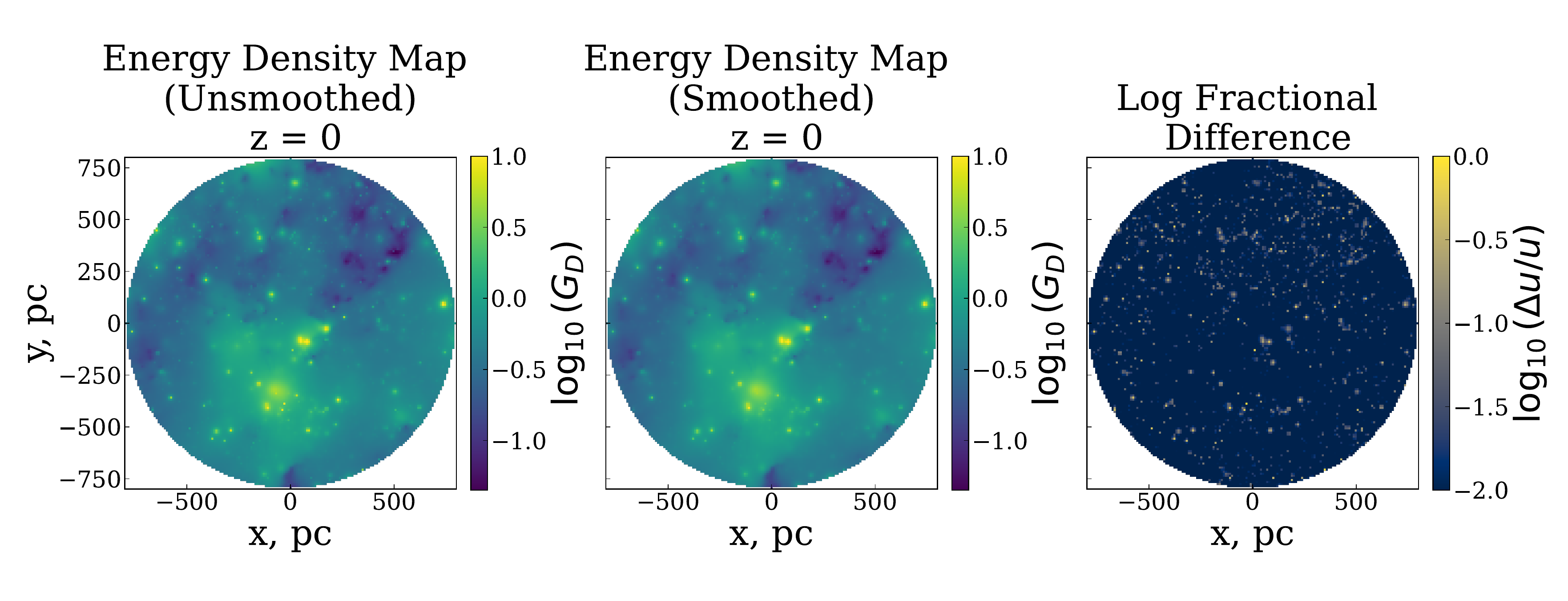}
    \caption{The left panel shows the slice of our standard map at $z = 0$, identical to the top right panel of Figure \ref{fig:dust_uv_0}.  The middle panel shows the same slice for the map in which the distances are smoothed using Equation \eqref{eq:smoothing}.  The right panel shows the pixel-by-pixel map of the fractional difference.  Most points are below the bottom limit of the color-bar, indicating a difference of less than 1\%.}
    \label{fig:smoothingPlot}
\end{figure*} 

{An unavoidable consequence of the finite grid spacing is that the value of $D_{ij}$ in Equation \eqref{eq:masterWithBins} may be a poor approximation for the typical distance from that star to positions within the voxel containing that star.}
Given that the median distance uncertainty for stars with $T_{\rm eff} >$ 10,000K within our sample is 12.78 pc (as compared with a cell-size of 10 pc), we do not believe that a careful averaging of the flux over a grid-cell volume is justified.  However, in some cases, the distance $D_{ij}$ is much less than 10 pc, and as a result a single star contributes disproportionately to the UV field within a given cell.  To investigate the magnitude of this effect, we constructed a map in which we replaced $D_{ij}$ in Equation \eqref{eq:masterWithBins} with a smoothed quantity
\begin{equation}
    D_{ij}^{\rm smoothed} = \sqrt{D_{ij}^2 + \frac{L_{\rm grid}^2}{12}},
    \label{eq:smoothing}
\end{equation}
where $L^2_{\rm grid}/12=8.33$ pc$^2$ is the variance of the distribution of the component of displacement from grid center along any of the three Cartesian axes. Equation \eqref{eq:smoothing} has the property that $D_{ij}^{\rm smoothed} \approx D_{ij}$ except in cases where the star in question lies within a particular grid cell, but limits the degree to which the contribution of a single star can be overestimated due to coincidental proximity to a grid-center. 

The results of this exercise are shown in Figure \ref{fig:smoothingPlot}.  The left and middle panels show slices at $z = 0$ from our standard map, and the one in which $D_{ij}$ is replaced with $D_{ij}^{\rm smoothed}$ as defined in Equation \eqref{eq:smoothing}.  The two maps are nearly indistinguishable by eye, though upon close inspection, one notes some pixels in which the UV field is much higher in the standard map.  The right panel shows a map of the fractional reduction in the FUV field that results from this smoothing.  
In this slice, differences greater than 1\% are present in only 9.1\% of the points, and differences greater than 10\% are present in only 0.69\% of points. 

\subsection{Related Concurrent Work}
In parallel work, Bish et al. (in prep) have used a different approach to make a similar map of the local UV field. In this paragraph, we outline the differences in the approach, and our perspective on the pros and cons of each. The present work uses stellar atmosphere models in combination with an algorithm to estimate the stellar type from optical/IR data. We account for dust extinction, but do not model the scattered radiation field. In contrast, Bish et al. (in prep) use observed UV fluxes from the TD1 catalog \citet{TD1} as sources in their model, and compute the resulting radiation field using a radiative transfer code that includes scattering. There are advantages and disadvantages to both approaches. The sophisticated radiative transfer modeling done in Bish et al. (in prep) allows for direct calculation of the scattered photon field, which we are only able to qualitatively discuss. Furthermore, by using directly measured UV fluxes, their method is less subject to uncertainties in the stellar parameters. Conversely, our model is not restricted to the discrete wavebands observed with TD1, and does not suffer from statistical noise due to propagating a finite number of photon packets. This approach allows us to use a substantially larger stellar sample.

\section{Conclusion}
\label{sect:conclusions}
We presented a map of the FUV energy density in the ISM within 800 pc of the Sun attributable to stellar continuum emission. We combined data from many sources to construct a catalog of stellar temperatures, sizes and locations.  We then used stellar models from \citet{Castelli03} to determine the UV luminosities and the 3D dustmap from \citet{Edenhofer24} to account for extinction by interstellar dust. All of the maps produced in this work, as well as the stellar catalog, are available at Harvard Dataverse \cite{Lomeli26_Data}.
\par
Near the Galactic midplane, we find a mean UV field of $5.3 \times 10^{{-14}}$ erg/cm$^3$, corresponding to 0.59 times the standard Draine field.  There is substantial spatial variation even within the midplane, with the 10$^{\rm th}$ percentile being a factor 2.54 below the mean, and the 90$^{\rm th}$ percentile a factor 1.78 above the mean. The mean UV field generally decreases with increasing gas density, except at gas densities (averaged over a cube of 10 pc on a side) of a few tens of hydrogen atoms per cm$^3$ which experience an elevated UV field as a result of their proximity to hot bright stars.  The average UV field generally decreases with increasing distance from the Galactic midplane, although this is likely due largely to the distance-limited nature of our stellar sample.
\\
\indent At the location of the Sun, we find an integrated energy density of ${6.56}\times10^{-14}$ erg/cm$^3$ (0.74 in units of the Draine field or 1.09 in units of the Mathis field). We compared this with observations from the Apollo 17 spacecraft, the TD-1 satellite, and the FIMS instrument.  We find that our model agrees with these observations to within a few tens of percent, depending on the wavelength range under consideration, except that we are missing the Lyman series emission lines seen in the FIMS data.  We note that our model does not have any tunable parameters, and therefore the broad agreement with observations suggest that we do not have any major systematic errors in our estimation of stellar UV luminosity.
\par
In addition to variations in the intensity of the UV field, we note some trends regarding the slope of the FUV spectrum.  First the UV spectrum is generally softer in high-density regions.  This trend is consistent with the fact that dust preferentially attenuates shorter-wavelength photons.  There is a strong positive correlation between the hardness of the spectrum and strength of the UV field, consistent with the (trivial) observation that regions of high UV field are preferentially near the hottest stars.  
\par
We discuss three sources of uncertainty in our modeling: omission of stars further than 1.25 kpc from the Sun, neglect of the scattered radiation field, and uncertainty due to the finite pixel size of our map. We estimate that the omission of stars beyond 1.25 kpc introduces an uncertainty of less than 10\% in most regions near the Galactic plane. Furthermore, we estimated the typical error incurred by neglecting the scattered radiation field to lie between 10 and 15\%, depending on wavelength. The magnitude of the estimated discrepancy generally increases with increasing gas density, increasing to $\sim 40\%$ in regions in which the gas density (averaged over our 10 pc pixel size) exceeds $n_H \approx 10$ cm$^{-3}$. To address the issue of finite grid resolution, we made a version of our map in which we artificially increased the distances of stars to grid centers to reduce any errors coming from stars whose positions are anomalously close to a particular cell center. We found that this made greater than a 10\% difference in less than 1\% of the pixels.
\\
{\it Acknowledgments:}  We gratefully acknowledge helpful discussions with Gordian Edenhofer. This work also benefited from helpful discussions with Hannah Bish and Joshua Peek at the Interstellar Institute 7 meeting at the Institut Pascal.
\software{Astropy \citep{astropy:2013, astropy:2018, astropy:2022}}

\bibliographystyle{aasjournal}
\bibliography{refs}

\appendix
\section{Derivation of Continuous Model Equations}
\label{sect:appendix}
Here we derive Equations \eqref{eq:Fcont} and \eqref{eq:taucont}.  We assume that the UV luminosity per unit volume $L$ and the differential UV extinction are Gaussian functions of of height $z$ above the Galactic midplane:
\begin{equation}
L= L_* \exp{\frac{-z^2}{2 \sigma_*^2}}; \quad \quad \tau' = \tau_0 \exp{\frac{-z^2}{2 \sigma_{\rm ext}^2}}.
\end{equation}
Then, the energy density at a point at a height $z$, only including stars out to distance R, is given by:
\begin{equation}
    u(z, R) = \frac{1}{4\pi c}\int^R_{r=0}2\pi rdr\int_{z-\sqrt{R^2-r^2}}^{z+\sqrt{R^2-r^2}}\frac{L_*\exp{\left(\frac{-z'^2}{2\sigma_*^2}\right)\exp{\left[-\tau_{\rm UV}(z', z, r)\right]}}}{r^2 + (z'-z)^2}dz.
\end{equation}
Here $\tau_{\rm UV}(z', z, r)$ is the optical depth in the UV between two points at height $z'$ and $z$ respectively, separated by cylindrical radius r. 
\begin{align}
    \tau(z', z, r) 
    &= \sqrt{1 + \left(\frac{|z' - z|}{r}\right)^2} \int_0^r 
    \tau'\!\left[z + (z' - z)\frac{r'}{r}\right]
    \, dr' \nonumber \\
    &= \tau_0 \sqrt{1 + \left(\frac{|z' - z|}{r}\right)^2}\int_0^r 
    \exp\!\left[-\frac{\left(z + (z' - z)\frac{r'}{r}\right)^2}{2\sigma_{\rm ext}^2}\right]
     \, dr'.
\end{align}

Let $r'/r=x$:

\begin{equation}
    \tau = \tau_0r\sqrt{1 + \left(\frac{|z' - z|}{r}\right)^2}\int_0^1\exp{\left[\frac{-[z + (z'-z)x]^2}{2\sigma_{\rm ext}^2}\right]}dx.
\end{equation}



Now, letting $Q = \frac{[z + (z'-z)x]}{\sigma_{\rm ext}\sqrt{2}}$, we find 
\begin{equation}
    \tau = \tau_0\sqrt{r^2 + |z'-z|^2}\int_y^w\frac{1}{w-y}e^{-Q^2}dQ,
\end{equation}
This then yields:

\begin{align}
    \tau = \sqrt{\frac{\pi}{2}}
    \frac{\sqrt{r^2 + |z-z'|^2}}{|z-z'|}
    \tau_0\sigma_{\rm ext}
    \left|
        \erf\left(\frac{z}{\sqrt{2}\sigma_{\rm ext}}\right)
        - \erf\left(\frac{z'}{\sqrt{2}\sigma_{\rm ext}}\right)
    \right|.
\end{align}

\section{Estimation of Stellar Radius}
\label{Stellar radius}

In order to estimate the radii of the stars in our catalog, we use two different methods based on the availability of parameters. One requires the availability of the Tycho ${V_T}$-band magnitude $m_{V_T}$, and the other makes use of the star's mass and log(g). If we can determine ${m_{V_T}}$, we solve for the radius that reproduces the correct ${m_{V_T}}$ given the estimated temperature, distance, and stellar extinction.  To do this, we use Equations 1 and 2 from \citet{Bessell}, substituting $(R/D)^2 e^{-\tau(\lambda)}{f_\lambda}$ in place of $f_\lambda$ in their Equation 2. This yields the relation


\begin{equation}
    {m_{V_T}}+2.5\log_{10}\left[\left(\frac{R}{D}\right)^2
    \frac{\int {f_\lambda} e^{-\tau(\lambda)} T_{V_T}(\lambda)\,\lambda\, d\lambda}
    {\int \frac{T_{V_T}(\lambda)c}{\lambda} d\lambda}\right] +48.6 - 0.037 = 0,
\label{eq:magnitudeToRadius}
\end{equation}
 where $R$ is the radius of the star, $D$ is the distance to the star, ${f_\lambda}$ is the stellar surface flux calculated with the models from \citet{Castelli03}, $T_{V_T}$ is the $V_T$-band transmission curve, $\tau(\lambda)$ the optical depth in wavelength $\lambda$ of the interstellar medium between us and the star, and -0.037 is the $V_T$ transmission filter's zero point. 
 \par
 Some stars do not have $V_T$ magnitudes from the previously mentioned catalogs. In the case that $V$ and $B$ or $G$, $B_P$, and $R_P$ magnitudes are available, we estimate a $V_T$ magnitude in the following way.\\
 \indent If a $V$ and $B$ magnitude are available, we use Equation 2.2.1 and 2.2.2 from \cite{esa_cosmos_sect2_02}:
 \begin{equation}
        V_T = V_{\text{mag}} + 0.09 \times \frac{(B_{\text{mag}} - V_{\text{mag}})}{0.85}.
\label{eq:V_to_VT}
    \end{equation}
Otherwise, if the Gaia $G$, $B_P$, and $R_P$ magnitudes are available, then we use the following equation from Table 5.6 in \cite{ESA_GEDR3_photRelations}:
\begin{equation}
V_T = G_{\text{mag}} + 0.01077 + 0.0682 \cdot (B_P - R_P)
+ 0.2387 \cdot (B_P - R_P)^2 - 0.02342 \cdot (B_P - R_P)^3.
\label{eq:VTfromGAIA}
\end{equation}
If we are unable to determine a $V_T$ magnitude, then we use a separate method. Starhorse-EDR3 catalog provides us with a value for log(g) and mass for each star.  In this case the stellar radius is given by
\begin{equation}
    R = \sqrt{GM/10^{\log(g)}}
    \label{eq:RfromMass}
\end{equation}
 where R, G and M are expressed in cgs units.

\end{document}